\documentclass[12pt]{aastex62}
\usepackage[utf8]{inputenc}
\usepackage[english]{babel}

\usepackage{amsmath}
\usepackage{amssymb}
\usepackage{amsfonts}
\makeatletter
\def\amsbb{\use@mathgroup \M@U \symAMSb}
\makeatother
\usepackage{bbold}
\usepackage{amsthm}
\usepackage{tabularx}
\usepackage[]{hyperref}
\usepackage{pbox}
\newtheorem{prop}{Proposition}
\theoremstyle{definition}
\newtheorem{defn}{Definition}

\DeclareMathAlphabet{\pazocal}{OMS}{zplm}{m}{n}
\DeclareMathOperator*{\argmin}{arg\,min}

\begin{document}
\title{A Framework for Telescope Schedulers: With Applications to the Large Synoptic Survey Telescope}

\author{Elahesadat Naghib}
\affil{Department of Operations Research and Financial Engineering, Princeton University, Princeton, NJ 08540, USA}
\email{enaghib@princeton.edu}

\author{Peter Yoachim} 
\affil{Department of Astronomy, University of Washington, Box 351580, U.W., Seattle, WA 98195, USA}
\email{yoachim@uw.edu}

\author{Robert J. Vanderbei} 
\affil{Department of Operations Research and Financial Engineering, Princeton University, Princeton, NJ 08540, USA}
\email{rvdb@princeton.edu}

\author{Andrew J. Connolly} 
\affil{Department of Astronomy, University of Washington, Box 351580, U.W., Seattle, WA 98195, USA}
\email{ajc26@uw.edu}

\author{R. Lynne Jones}
\affil{Department of Astronomy, University of Washington, Box 351580, U.W., Seattle, WA 98195, USA}
\email{ljones@astro.washington.edu}


\begin{abstract}
How ground-based telescopes schedule their observations in response to competing science priorities and constraints, variations in the weather, and the visibility of a particular part of the sky can significantly impact their efficiency. In this paper we introduce the Feature-Based telescope scheduler that is an automated, proposal-free decision making algorithm that offers \textit{controllability} of the behavior, \textit{adjustability} of the mission, and quick \textit{recoverability} from interruptions for large ground-based telescopes. By framing this scheduler in the context of a coherent mathematical model the functionality and performance of the algorithm is simple to interpret and adapt to a broad range of astronomical applications. This paper presents a generic version of the Feature-Based scheduler, with minimal manual tailoring, to demonstrate its potential and flexibility as a foundation for large ground-based telescope schedulers which can later be adjusted for other instruments. In addition, a modified version of the Feature-Based scheduler for the Large Synoptic Survey Telescope (LSST) is introduced and compared to previous LSST scheduler simulations.
\end{abstract}


\section{Introduction}
The Large Synoptic Survey Telescope (LSST), is a large,  ground-based optical survey that will image half of the sky every few nights from Cerro Pachon in Northern Chile. The LSST comprises an 8.4 meter  primary mirror and a 3.2 Gigapixel camera.  With a 9.6 $\text{degree}^2$ field-of-view, it will visit each part of its 18000 $\text{degree}^2$ primary survey area ~1000 times over the course of 10 years. Each visit will likely comprise a 15 second pair of exposures with a single visit depth of ~24.5 magnitudes (AB) (in the six bands u, g, r, i, z, and y). The revolutionary role of this telescope calls for no less than optimal operation. 

The algorithm that makes the sequential decisions of which filter to use, and which direction to point the telescope to, is called \textit{scheduler}. A scheduler has to maximize the scientific outcome of the telescope during its limited period of operation. 

There are four primary science drivers for the LSST project: the characterization of dark energy through the multiple cosmological probes (e.g. gravitational weak lensing, luminosity distances from Type Ia supernovae, and Baryon Acoustic Oscillations); mapping the 3D distribution of stars within our Galaxy; a census of solar system objects within the Solar System; and a detailed study of the transient and variable universe. Each of these objectives has a different set of constraints and requirements on how the observations are made (e.g. the cadence of the observations, the number of filters as a function of time, the acceptable airmass range for an observation). The mission of modern, large, ground-based telescopes, such as LSST, is constrained with various stochastic factors, and contains competing objectives which are vastly different in nature, due to the different nature of the scientific expectations. In this paper, we propose a framework to formulate the problem of scheduling for the new generation of ground-based telescopes then introduce a scheduler based on the proposed model.

The first generation of schedulers for astronomical instruments were developed for space missions mainly to automate their operation. ROSAT mission's scheduler in 1990 \citep{nowakovski1999using},  Spike \citep{johnston1994spike}, Hubble Telescope's scheduler in 1994, and HSTS \citep{muscettola1995automating} in 1995 pioneered many of the developments in algorithmic scheduling of observations for the space missions.

Despite the similarity of the science objectives for space and ground-based telescopes, the determining factors for the purpose of scheduling are fundamentally different. While space telescopes are required to respect kinematic and dynamical constraints, weather is the main challenge in the scheduling of ground-based telescopes. The former is predictable and efficiently computable, while the later involves both inherent uncertainties, and uncertainties due to computational limitations.

Earlier algorithmic approaches to the scheduling of ground-based telescopes are heavily based on observation \textit{proposals} which are hand-crafted sequences of scripted astronomical observations. Proposals are generally tested only for feasibility (e.g. that a set of fields were visible, or lay within a specified airmass range, or within a window in time), but not necessarily for optimality. For instance, the operation of Keck Telescope, 1993 \citep{nelson1985design}, is 100\% based on proposals, while the Hobby-Eberly Telescope, 1997 \citep{shetrone2007ten}, has a semi-manual scheduling scheme.

More recently, the development of more expensive ground-based instruments with complex missions made it impossible to rely solely on hand-crafted proposals. The need for more efficient use of the instrument's time led to the development of decision making algorithms to optimize science output. For instance, the scheduler of the Liverpool robotic telescope was designed in 1997 to automate and optimize time allocation into chunks of scripted schedules. The time allocation strategy was preferred to the scheduling at the single visit level. The latter approach is referred to as optimum scheduling by the authors \citep{steele1997control}. The reason for choosing the time allocation scheme instead of optimum scheduling is stated to be the lack of recoverability of the latter choice in case of an interruption, because it potentially leads to excessive computational cost to reevaluate the sequence once it is interrupted. In this paper, we show that the scheduling in the single visit level can be quickly recovered in a memoryless framework, thus the optimality is not necessarily required to be sacrificed. Another example is the Las Cumbres Observatory Global Telescope Network (LCOGT) with one of the most advanced telescope scheduling algorithms \citep{Boroson14, Saunders14}. LCOGT uses an integer linear programming (ILP) model, solved with the Gurobi algorithm \citep{gurobi}, to optimize the scheduling of observations over a global network of telescopes \citep{Lampoudi15}. Due to the success of this approach the Zwicky Transient Facility at Palomar Observatory \citep{Bellm14} has also adopted an ILP scheduler\footnote{GitHub repository: \url{https://github.com/ZwickyTransientFacility/ztf_sim}}. The ILP scheduling model performs well for observatories where slew overheads are small compared to exposure execution time. LCOGT is able to schedule observations in long contiguous blocks. In contrast, the LSST plans to have observations of about 30 seconds with slew times up to 2 minutes (when there's a filter change), hence it requires a scheduling algorithm that can explicitly minimize the slew times between successive observations. One could use the ILP scheduling approach to schedule large prescript blocks of observations for LSST. The blocks could be set to follow a path that only includes short slews (this is similar to the strategy taken in the scheduler developed by D. Rothchild et al.). The disadvantage of this approach and any other prescript schedule is principally the lack of recoverability from unpredictable conditions, for instance inability to dodge the clouds.

Even given the reliability of fully automatic scheduling technologies, remain a number of modern telescopes such as SALT, 2005 \citep{brink2008salt}, and ALMA, 2013 \citep{wootten2003atacama}, which are being operated based only on traditional hand-crafted proposals. ALMA in particular, requires a highly regulated structure for proposals that potentially leads to suboptimality as discussed in \citep{alexander2017enabling}, where the authors suggest a number of corrections for the scheduling regulations to provide adaptivity to time-sensitive observations.

The LSST community, however, in addition to a proposal-based scheduler, introduced in \citep{delgado2016lsst}, have been supporting the design and implementation of proposal-free decision machines, such as Feature-Based scheduler, first introduced in \citep{naghib2016feature}, as well as a semi-scripted cadence by D. Rothchild et al. (2018, in preparation) that explore the possibility of a new generation of the schedulers for fast, multi-mission, big-data collecting instruments.

In Section~\ref{sec_SM}, first we explain the choice of Markovian framework for the Feature-Based scheduler, and in Sections~\ref{sec_Markov} and~\ref{sec_Markov_approx} we provide the mathematical details of the scheduler model in that framework. Section~\ref{sec_opt} presents two approaches for the optimization of the model's parameters. Sections~\ref{sec_lsst_problem} demonstrates the application of the Feature-Based scheduler on the LSST which is then followed by a comparison between a modified version of the Feature-Based scheduler and LSST's current official scheduler in Section~\ref{sec_comp}. Finally, Section~\ref{sec_conclusion} presents our concluding remarks.

\section{Scheduling framework}\label{sec_SM}
To run a ground-based telescope with multiple science objectives, such as LSST, the scheduler has to offer \textit{controllability}, \textit{adjustability}, and \textit{recoverability}.
\begin{itemize}
\item \textbf{Controllability}: The high-level science objectives have to appropriately respond to the variation of the model parameters. Otherwise, either the choice of the input information is irrelevant to the mission objectives, or the structure of the scheduler is degenerate. Controllability is necessary for a scheduler to be optimizable by the choice of the model parameters. 
\item \textbf{Adjustability}: For a complex, multi-objective mission it is common that the high-level objectives are required to be modified in the middle of the operational period. Adjustments take place according to updates of scientific goals, or changes in the mechanical performance of the system. Regardless of the reason for the adjustments, a scheduler must offer flexibility to be adjusted with a reasonable computational cost, and preferably no expert intervention. Hand-tuned scheduling strategies and verbal policies for instance, are not fully adjustable.

\item \textbf{Recoverability}: The presence of unpredictable factors in the operation of ground-based telescopes are due to the natural stochastic processes (such as the weather), and complexity of the mechanical facility. Unscheduled downtime and instrument failures are examples of the many unpredictable survey interruptions. On the other hand, there are inherently predictable interruptions, such as maintenance downtimes, and cable winding that also, due to the complexity of the mechanical system, are not computationally affordable and/or valuable to keep track of. Therefore, they are considered as stochastic variables as well. Given all the stochastic factors, a scheduler is required to be able to quickly make an alternative decision, once a previously unpredictable event occurred. A scripted sequence of decisions for instance, lacks the recoverability attribute. Also, strategies that need to look back at historical sequences of events or look forward through possible sequences of events are not fast recoverables. 
\end{itemize}

The Feature-Based scheduler is designed based on a Markovian model in which the flow of the input information, the decision procedure and it's relationship to the mission's objective are coherently expressed. Therefore, controllability of the scheduler is well-defined and verifiable. And it is adjustable, because of the Markovian structure that offers an explicit derivation of the design elements from high-level objectives, and finally is swiftly recoverable due to the inherent memorylessness of the Markovian Decision Process which for a decision at any time, only requires the current state of the system.

\subsection{Markovian representation}\label{sec_Markov}
\begin{defn}
Let $X_{(.)}$ be a stochastic process for which $X_i$ represents the state of the system at $t_i$, and $\pazocal{S}$ be the set of all possible states that the system can take. Let $\mathbb{P}(X_i)$, be the probability distribution of $X_i$ on $\pazocal{S}$, then $X(.)$ is a Markovian process, if and only if, it satisfies the following \textit{memorylessness} property,
\begin{equation*}
\forall i ~~~~~~~\mathbb{P}(X_{{i+1}} | X_{i}) = \mathbb{P}(X_{{i+1}} | X_{i}, X_{{i-1}},\dots, X_{0}),
\end{equation*}
where $\mathbb{P}(X_{{i+1}} | X_{i})$ is the conditional probability distribution of the system's state at $t_{i+1}$ given it's state at $t_i$, and $\mathbb{P}(X_{{i+1}} | X_{i}, X_{{i-1}},\dots, X_{0})$ is the conditional probability distribution of the system's state at $t_{i+1}$, given all of the states that the system has been in until $t_i$.
\end{defn}
Memorylessness property asserts that the system's next state only depends on its current state and is independent of its earlier history. This property, is in fact, the main reason for choosing a Markovian framework for the scheduler. 
\begin{defn}
Let $<\pazocal{S},\pazocal{A}, P_a(.,.), R_a(.,.), \gamma>$, be a \textit{Markovian Decision Process (MDP)}, where $\pazocal{A}$ is the set of actions, and $P_a(x, y)$ transition probability from state $x$ to $y$ which is equal to $\mathbb{P}(X_{i+1}=y | X_i=x, a)$, the conditional probability of transition from state $x$ to state $y$ given action $a \in  \pazocal{A}$. Finally the transition reward is denoted by $R_a(x,y)$, and $\gamma \in (0,1]$ is the discount factor.
\end{defn}
\begin{defn}
Action $a_{i} \in \pazocal{A}$ is \textit{admissible} for $<\pazocal{S},\pazocal{A}, P_a(.,.), R_a(.,.), \gamma>$, if it is \textit{feasible}, thus it is possible to be taken at $t_i$, and \textit{progressively measurable}, hence only dependent on the current state of the system $X_i$.
\end{defn}
To control the system is to take an action at all decision steps $t_i$. For a Markovian control, the actions are required to be dependent only on the current state to preserve the memorylessness property of the closed-loop system, that's why we require the action to be progressively measurable. Notice that the decision steps $t_i \in \{ t_j: 0 \leq j \leq N, t^N  =T\}$, are not necessarily uniformly spaced, and are determined by the time that each transition takes. In this representation, $t_0$ is the start of the process and $T$ is the finite time horizon of the process.

\begin{defn}
A \textit{deterministic policy} $\pi : \pazocal{S} \rightarrow \pazocal{A}_i$, is a mapping from $\pazocal{S}$ to the set of all admissible actions at $t_i$, denoted by $\pazocal{A}_i$. 
\end{defn}
A policy provides a time invariant control law that for all possible $x_i \in \pazocal{S}$ suggests an admissible action for transition to the next state, which automates the control of the system.
\begin{defn}
\textit{A deterministic optimal policy} $\pi$ is a solution to the following optimization problem,
\begin{equation}\label{equ_opt1}
\begin{aligned}
& \underset{\pi}{\text{maximize}}
& & E_{\pi}[\sum_{i = 0}^N \gamma^i R_{\pi(X_{i-1})}(X_{i-1}, X_{i}) | x_0],
\end{aligned}
\end{equation}
where $x_0$ is a given initial state.
\end{defn}
In other words, a deterministic optimal policy maximizes the expected discounted sum of the rewards. By discounting the later rewards versus earlier rewards through $0 < \gamma \leq 1$, we tune the priority of the overall gain versus instant gains.
\begin{prop} \label{prop_main}
For the Markov decision process of $<\pazocal{S},\pazocal{A}, P_a(.,.), R_a(.,.), \gamma>$, there exists a deterministic optimal policy, and it can be written as follows, 
\begin{equation}\label{equ_opt_pol}
\pi^*= \argmin_{a_{i} \in \pazocal{A}_i} E[\Phi(X_{{i+1}}) | a_{i}],
\end{equation}
where $\Phi: \pazocal{S} \rightarrow \rm I\!R$ is a function of the following form,

\begin{equation}\label{equ_phi_compatible}
\Phi(x_{i}) =  - R_{\pi^*(x_{i-1})}(x_{i-1},x_i)  + \gamma E_{\pi^*}[\Phi(X_{{i+1}})|x_i].
\end{equation}
\end{prop}

\textit{Proof.} See Appendix.

For the telescope scheduler we require the policy to be deterministic, because the simulations have to be repeatable for comparison and evaluation purposes, however it can be shown that the deterministic optimal policy is not only optimal amongst deterministic policies, but also is optimal amongst stochastic policies. Therefore the choice of deterministic policy does not harm the optimality of the control. 

As a result of Proposition~\ref{prop_main}, the solution space of the optimization problem (\ref{equ_opt1}), can be reduced from search over the set of policies (all possible mappings) to search over $\Phi$ functions.

\subsection{Markovian approximation}\label{sec_Markov_approx}
For a decision that is inherently time dependent, such as scheduling an observation, only a maximal definition of the system's state yields a perfect Markovian system, in which case, the state-space includes all of the possible decision sequences. In particular, LSST requires a sequence of about 1000 decisions at each night. Therefore, storing all possible scenarios requires a state space of size $N_{f}^{1000}$, where $N_f$ is the number of possible printings on the visible sky. No matter how one tessellates the sky,  $N_{f}^{1000}$ number of scenarios is neither tractable nor storable in a realistic memory. In order to overcome the curse of dimensionality, we have designed a set of   \textit{features} to summarize the most important information required for the scheduling. Thus, the system of telescope-environment is only an approximated Markovian system, once its state-space is replaced by a \textit{feature-space}. 

On the other hand, Proposition (\ref{equ_opt1}) shows that the optimal scheduler lies within the set of functions instead of the much larger set of all possible mappings. Despite this reduction, Problem (\ref{equ_opt1}) is still an infinite dimensional optimization problem because its variable is a function. To be able to numerically compute the $\Phi$ function, we propose a parametrized function approximation for $\Phi(x_{i})$,
\begin{equation*}
 \tilde{\Phi}_{\theta}({x}_{i}) := \sum_{j=1}^m \theta_j \Phi_j(x_{i}),
\end{equation*}
where $\theta$ is the vector of variables that characterize $\tilde{\Phi}(.)$, and $\Phi_i(x_i)$'s are the \textit{basis functions} that are designed to modify the features in order to incorporate the astronomical observational knowledge into the decision maker's structure. With this approximation, the space of search is reduced from the space of functions to a finite dimensional vector space. This approximation substitutes the original optimal policy (\ref{equ_opt_pol}) with the following approximate policy,
\begin{equation}\label{equ_approx_pol}
\begin{aligned}
\tilde{\pi}_{\theta}^*(x_{i})& = \argmin_{a_{i}} E[ \tilde{\Phi}_{\theta^*}(X_{i+1}) | a_{i}] =  \argmin_{a_{i}} \sum_{j=1}^k \theta^*_j E[\Phi_j(X_{i+1}) | a_{i}],\\
\end{aligned}
\end{equation}
where $\theta^*$ is a solution to the following optimization problem, in which policy $\pi$ is fully determined by $\theta$.
\begin{equation}\label{equ_opt3}
\begin{aligned}
& \underset{\theta}{\text{maximize}}
& & E_{\pi_{\theta}}[\sum_{i=0}^N \gamma^i R_{\pi_{\theta} (X_{i-1})}(X_{{i-1}}, X_{i}) | x_0],
\end{aligned}
\end{equation}

\section{Scheduler optimization}\label{sec_opt}
Given the approximated optimal policy in Equation (\ref{equ_approx_pol}), the only remaining fundamental step to have a scheduler, is to find $\theta^*$ by solving Problem (\ref{equ_opt3}). The following two sections describe two different approaches to find a $\theta^*$. The first approach requires a specific class of the high-level mission objective function, and is faster. The second approach is applicable for all types of the high-level mission objective functions, however requires more computational resources. 

\subsection{Reinforcement Learning}

Assume that there exists a well-defined notion of the instant reward for each state transition, then $\Phi(x_{{i}})$ by definition given in Equation (\ref{equ_phi_compatible}) is,

\begin{equation} \label{equ_learning1}
\Phi(x_{i}) = - R_{\pi^*(x_{i-1})}(x_{i-1},x_i)  + \gamma E_{\pi^*}[\Phi(X_{{i+1}})|x_i].
\end{equation}

Accordingly, for the parameterized $\Phi$ function, we require the following,

\begin{equation} \label{equ_learning2}
\begin{aligned}
\tilde \Phi_{\theta^*}(x_{i}) &=  - R_{\pi_{\theta^*}(x_{i-1})}(x_{i-1},x_i)  + \gamma E_{\pi_{\theta^*}}[\tilde \Phi_{\theta^*}(X_{{i+1}})|x_i].\\
\end{aligned}
\end{equation}

In reinforcement learning, the main idea to find $\tilde {\Phi}_{\theta^*}(.)$ is to start the process of the decision making with an arbitrary set of variables, $\theta^0$, and make the decisions according to Policy (\ref{equ_approx_pol}) with the associated $\tilde \Phi_{\theta}(.)$, then update the variables in each decision step, so that $\Phi$'s linear approximation gradually respects Equation (\ref{equ_learning2}) for all $i \in \{j: t_0 \leq t^j \leq T\}$.

Note that at $t_i$, after the transition from $x_{{i-1}}$ to $x_{{i}}$ we have the value of $R_{\pi_{\theta^i}}(x_{i-1},x_i) $ already evaluated in the decision making procedure, and  $\Phi_{\theta^i}(x_{i})$ can be approximated by $E_{\pi_{\theta^i}}[ \Phi_{\theta^i}(x_{i})]$ which is also evaluated in the decision making process, where $\theta^i$ is the last version of the optimization variables at $t_i$. Using the desired value given in Equation (\ref{equ_learning2}), the update is as follows,

\begin{equation}\label{equ_updatePHI}
\begin{aligned}
\tilde \Phi_{\theta^{i+1}}(x_{{i}}) = (1-\alpha) \tilde \Phi_{\theta^{i}}(x_{{i}})+ \alpha  ( \gamma  E_{\pi_{\theta^i}}[\tilde \Phi_{\theta^{i}}(x_{{i+1}})|\pi_{\theta^{i}}(x_{{i}})] -R_{\pi_{\theta^i}}(x_{i-1},x_i) ),
\end{aligned}
\end{equation}
in which, $0<\alpha<1$ is the learning rate. The first term in the right hand side of the equation is the latest approximated value of the $\tilde \Phi$ function associated with $\theta^i$ with $(1-\alpha)$ amount of contribution, and the second term is the value of $\tilde \Phi$ function according to Equation (\ref{equ_learning2}) with $\alpha$ amount of contribution. Clearly for smaller $\alpha$'s this update imposes smaller adjustments. Accordingly, updates of the variables $\theta_j,~~j=1,\dots, k$ can be expressed as follows,

\begin{equation} \label{equ_TD_update}
\begin{aligned}
\theta_j^{i+1} &= \theta_j^{i} + \Big( \tilde \Phi_{\theta^{i+1}}(x_{{i}})  - \tilde \Phi_{\theta^{i}}(x_{{i}}) \Big)\Phi_j(x_{i})\\
& = \theta_j^{i} + \alpha \Big(-\Phi_{\theta^{i}}(x_{{i}}) + \gamma  E_{\pi_{\theta^i}}[\tilde \Phi_{\theta^{i}}(X_{{i+1}})|x_{{i}}] -R_{\pi}(x_{i-1},x_i)  \Big)\Phi_j(x_{i}).
\end{aligned}
\end{equation}

This Learning method is called \textit{Temporal-Difference} (TD) learning with function approximation \citep{tsitsiklis1997analysis}. Variants of this reinforcement learning method have been successfully applied to real-life problems such as training of a backgammon player \citep{tesauro1995temporal}.

Note that to be able to use the TD reinforcement learning method, it is necessary to have a well-behaved notion of the reward that reflects the instant gain of any decision at all of the decision steps. Moreover, the discounted sum of the instant rewards has to reflect the objective of the mission. For instance, in the LSST scheduling problem, after each visit, the negative of the slew time, is a well-defined instant reward that reflects how time-efficiently the telescope is being used. This however does not reflect all aspects of the mission's objective such as, the need to re-observe a field within a valid time window (explained in Section~\ref{sec_lsst_problem}), for which there is no equivalent instant reward. For this reason, we also implemented a black-box function optimizer for the LSST scheduler that does not require a notion of the instant reward and directly optimizes the mission's objective function over a limited  episode of the simulated scheduling. 

\subsection{Global Optimization}\label{sec_gopt}
In the absence of a well-defined instant reward, instead of solving problem~(\ref{equ_opt3}), The following problem can be solved,

\begin{equation}
\begin{aligned}
& \underset{\theta}{\text{maximize}}
& & U_{\pi_{\theta}}(x_i,x_{i+1}, \dots, x_{j}),
\end{aligned}
\end{equation}
where $U_{\pi_{\theta}}(x_i,x_{i+1}, \dots, x_{j})$ is a utility function that measures the performance of the scheduler on a simulated episode of the operation from $t_i$ to $t_j$ by policy $\pi_{\theta}$. In this approach, a general $U(.)$, can not be explicitly expressed as a function of $\theta$, therefore a global optimizer that can maximize a black-box function is required. Evolutionary optimizers have successfully been applied to numerous real-life problems involving black-box function optimization, and specifically astronomical mission planning such as the scheduling of Exoplanet Characterisation Observatory \citep{garcia2015artificial}. We used the $e$DE evolutionary optimizer \citep{naghib2016entropic} which is an adaptive version of the Differential Evolution (DE) algorithm (\citep{storn1997differential}). DE is generally one of the most efficient evolutionary algorithms and the $e$DE variant uses a notion of entropy to automatically preserve the diversity of the candidate solutions. As a result, in contrast with DE, it does not require the user to choose any tuning parameters for the algorithm, which is the most time-consuming task in using an evolutionary optimizer. In addition, $e$DE, similar to any other evolutionary algorithm is highly parallelizable, and the computational time can be almost linearly decreased with respect to the number of computational cores.

\section{Problem of Scheduling for LSST}\label{sec_lsst_problem}

\begin{table}[h]
\caption{Key terms and notations used in the definition of the features and the basis functions} \label{tab_notatopn}
\begin{tabular}{l  l }
\hline
fields & fixed point configuration on the sky such that the visible sky could be completely covered by pointing\\
& the telescope toward all of those directions, \\
$N_{f}$ & total number of the fields,\\
$t$& Coordinated Universal Time (UTC),\\ 
$\tau_s(t)$ ($\tau_e(t)$)& beginning (end) of the night that $t$ lies within,\\
$\tau_{rise\text{(}set\text{)}}(i,f,t)$ & rising (setting) time of field-filter $(i,f)$ above (below) the acceptable airmass horizon at current night,\\
$\tau_n(i,f,t)$& time of the last visit of field-filter $(i,f)$ before $\tau_s(t)$,\\
$id(t)$ & ID number of the field that is visited at $t$,\\
$ft(t)$& camera's filter at $t$,\\
$n(i,f,t)$ & total number of the visits of field-filter $(i,f)$ before $t$,\\
$slew(i,j)$& slew time from field $i$ to field $j$ in seconds,\\
$settling(i,j)$& mechanical settling time after slewing from field $i$ to field $j$,\\
$\Delta t_{f}$& time needed to change filter, a constant value about 2 minutes,\\
$t_{dome}(i)$& time needed to move the dome to make field $i$ visible to the telescope,\\
$ha(i,t)$ & hour angle of the center of field $i$ at $t$ in hours, $-12 \leq ha(i,t) \leq 12$,\\
$am(i,t)$ & airmass of the center of field $i$ at $t$,\\
$br(i,t)$ & brightness of the sky at the center of field $i$ at $t$,\\
$\sigma(i,t)$ & seeing of the sky at the center of field $i$ at $t$,\\
$K(i,f, t)$ & atmospheric extinction coefficient\\
$W_1, W_2$ & given constant time window within which a revisit is valid, \\
\hline
\end{tabular}
\end{table}

The LSST's mission is to uniformly scan the visible sky within 5 different regions shown in Figure~\ref{fig_proposals}. Each region, also referred to as \textit{survey}, has certain science-driven goals and constraints, defined and precisely described in \citep{ivezic2008large}.  

\begin{figure}[h!]
\begin{center}
\epsscale{0.5}
\plotone{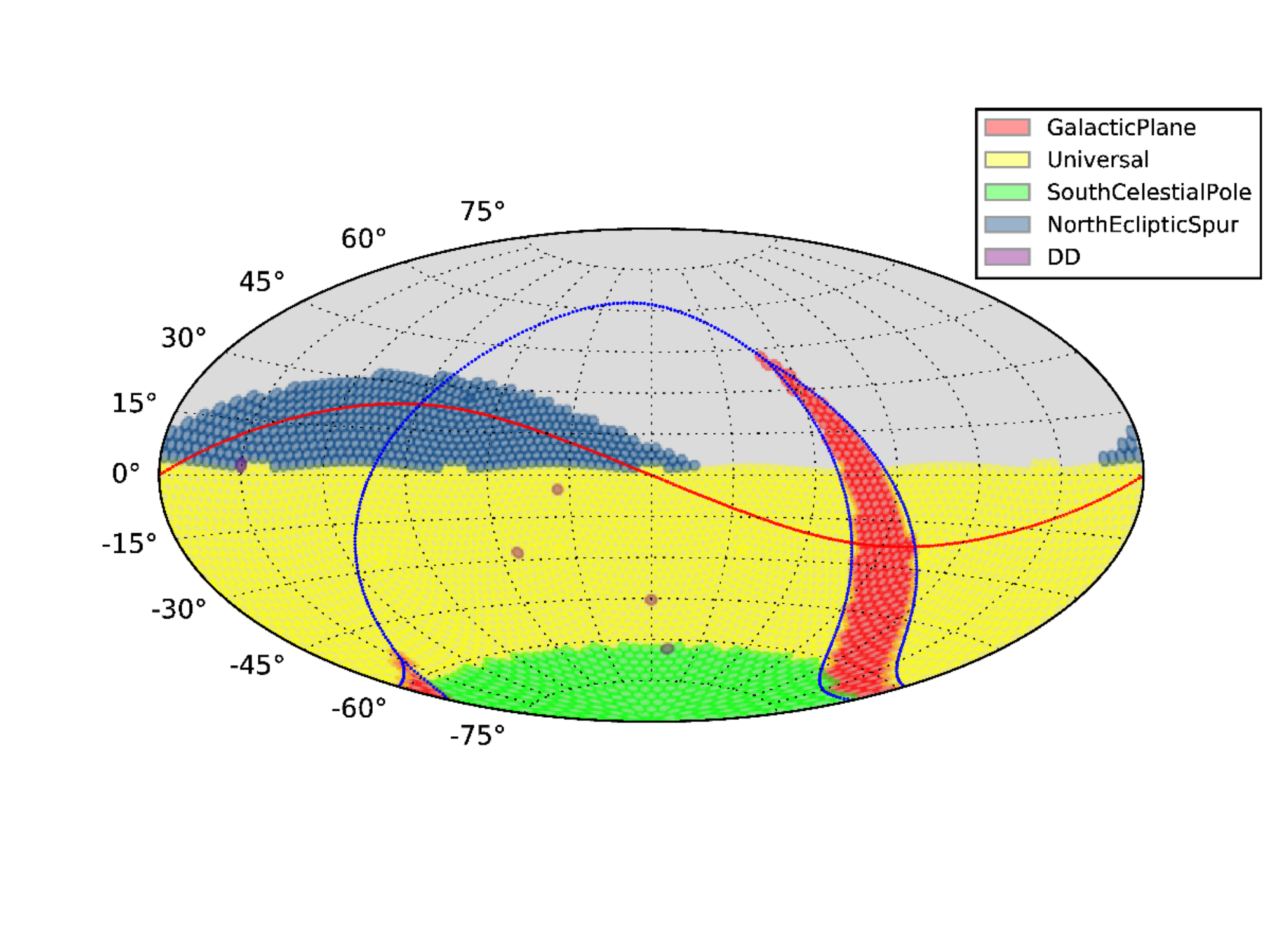}
\caption{Regions of the sky with different requirements and constraints for scheduling: (1) Galactic Plane Region (GP), (2) Universal or Wide Fast Deep (WFD), (3) South Celestial Pole (SCP), (4) North Ecliptic Spur (NES), and (5) Deep Drilling Fields (DD).}
\end{center}
\end{figure}\label{fig_proposals}

 The notion of the features, enables the scheduler to systematically fetch all of the various requirements and turn them into comparable quantities for the purpose of the decision making in the Markovian framework. The proposed feature-space of the LSST contains seven features, each can be evaluated given a field $i$, a filter $f$, and a time $t$. The fields discretize the visible sky through a fixed partitioning. Each field can be captured through a single visit, and there are 6 possible filters, $[u,g,r,i,z,y]$, for each visit. And finally the time domain is discretized by the natural timing of the process. In other words, time intervals at which you have to make a decision are heterogeneous. Given that a consecutive visit of the same field-filter is not allowed in the main survey, there is a slew time between any two decisions, therefore $t_{j} - t_{j-1} > 0$. On the other hand, the operation is over a limited time horizon, $T$, thus the number of the decision time steps is finite. In conclusion, a finitely discretized sky, a finite number of the filters and a finite number of the time steps, pose a finite feature-space, denoted by $\{(f_1(i,f,t_j)\dots f_7(i,f,t_j)): i = 1\dots n_f, f\in \{u,g,r,i,z,y\}, j = 0,\dots N\}$. Then the implication of the policy, stated in Equation (\ref{equ_approx_pol}), for the LSST scheduler would be as follows,

\begin{equation}\label{equ_approx_pol_imp}
\tilde{\pi}^*(x_j)= \argmin_{(i,f)\in \pazocal{A}_{j}} \sum_{k=1}^5 \theta^*_k E_{\pi}[\Phi_k(X_{j+1}) | x_j]\\,
\end{equation}
where $x_j = [f_{1, \dots,7}(id(t_j),ft(t_j),t_j)]$ is the 7-dimensional state at $t_j$, and $(i,f)$ is a feasible pair of field-filter. Section~\ref{sec_cstr} introduces the constraints under which a field-filter pair, $(i,f)$ is feasible at $t_{j}$. Accordingly, $\pazocal{A}_j$ is the set of all field-filter pairs that are feasible at $t_j$. Table ~\ref{tab_feasibility} contains the feasibility conditions.

For a modular approach to the implementation of the scheduler, expected values of the basis functions, $E_{\pi}[\Phi_k(x_{j+1})|x_j]$ for $k=1,\dots, 5$, are evaluated in separate modules. Those basis functions that address the environmental parameters are developed by the LSST community. See \citep{2014SPIE.9145E..1AG} for the parameters capture the status of the LSST site, \citep{sebag2008lsst} and \citep{sebag2007lsst} for cloud cover measurements that were used to develop a cloud model and, see \citep{yoachim2016optical} for the sky brightness model. Generally speaking, making a decision for a visit at $t$ for the LSST scheduling problem is mainly determined by the following factors,

\begin{enumerate}
\item The amount of the time it takes to redirect the telescope and the dome to move from one target to the next target.
\item The short-term science-driven requirements, such as the same-night revisit of a field.
\item The long-term mission-driven requirements, such as maintaining a uniform coverage of all field-filter pairs within each region.
\item The relative quality of visiting a field-filter compared to other field-filters at the time of observation, for overall efficiency of the operation.
\item The general preference for observing the fields around the meridian.
\end{enumerate}

Accordingly, the basis functions of the LSST scheduler, are designed to formalize the above factors. For the full definition and description of the basis functions of the Feature-Based scheduler for LSST refer to Section~\ref{sec_lsst_bfs}.

The last step is to implement the training procedures (described in Section~\ref{sec_opt}) on the LSST model to derive $\theta^*$. Training of the LSST scheduler is explained in Section~\ref{sec_lsst_opt} with two sample objective functions. However, the LSST community, and basically any individual can design their own mission objective function, whether it allows for well-defined instant rewards or not, then train the scheduler through our open source training code, and find a new set of $\theta^*$ that principally leads to a different behavior of the scheduler due to different objectives.

\subsection{Features of the telescope scheduler}\label{sec_lsst_features}

For designing the features, it is important to avoid redundancy in the information that features contain. It is also critical to hold a modular approach in the delivery of the information to the decision procedure. For instance, consider the amount of time, $\Delta t$, it takes for a telescope to move on from a visit. In the LSST problem, $\Delta t$ mainly depends on the slew time, mechanical settling time, the dome placement time, and the time it takes to change the filter. All of these timings are available through a precise simulation of the LSST model \citep{delgado2014lsst}. A modular design would be to bring the summation of the operational timings to the stage of the decision, instead of bringing them separately as different features. This approach makes the implementation significantly simpler, and more readable, and in a conceptual level, makes it possible to track the effect of the operational timing in the overall outcome of the decision maker. Particularly, since the operational cost is independent of the amount to which each cause contributes to the overall $\Delta t$, bringing the timing of the separate procedures separately in the decision making level adds unnecessary complications to the design, and consequently, makes it hard to back track the output-input behavior of the scheduler. 

This section proposes seven features for the description of the LSST-environment state in Table \ref{tab_features}, with the key terms and notations defined in Table \ref{tab_notatopn}. They are designed to efficiently carry the determining information with a modular approach. Each feature is denoted by $f_k(i,f,t)$ for $k= 1..7$, and indexed by the triplet of $(i,f,t)$, field, filter, and time. To make a decision at $t$, the scheduler computes principally all of the seven features for all of the $(i,f)$ pairs, however, there are some features that don't change in every time step, for instance, if $i$ is not visited at $t^j$, then $f_5(i,f,t^j)= f_5(i,f,t^{j+1})$. For such cases, the implementation has a categorized updating procedure to avoid redundant computations.

\begin{table}
\caption{Features of the approximated Markovian model for the telescope-environment system. Features $f_{1\dots 7}$ provide a memoryless, approximate description of the system's state.}
\begin{tabular}{|l|l|}
\hline
Notation & Definition\textbackslash Description\\ \hline \hline 
$f_1(i,f,t)$ & \pbox{0.85\textwidth}{($slew(id(t),i)+settling(id(t),i)+\Delta t_{f} I_{ft(t) \neq f} \vee t_{dome}(i)$): either the time required to point the telescope to $i$, and change the filter to $f$, or the time required to relocate the dome to make $i$ visible. Whichever that is larger.}\\ \hline
$f_2(i,f,t)$ & \pbox{0.86\textwidth}{the total number of the same-night visits of field-filter $(i,f)$ until $t$.}\\ \hline
$f_3(i,f,t)$ &  \pbox{0.86\textwidth}{$(t - \tau_n(i,f,t)) I_{\{\theta(i,f,t) > \tau_s(t)\}}$, time since the last same-night visit of $(f,i).$}\\ \hline
$f_4(i,f,t)$ &  \pbox{0.86\textwidth}{remaining time for field-filter $(i,f)$ to become invisible, either by passing the airmass or the moon-separation limit, or being covered by temporary objects such as clouds, as projected at $t$.}\\ \hline
$f_5(i,f,t)$ &  \pbox{0.86\textwidth}{co-added depth, a measure of cumulative quality of past visits of field-filter$(i,f)$ until $t$.}\\ \hline
$f_6(i,f,t)$ &  \pbox{0.86\textwidth}{$5\sigma$-depth, a measure for quality of visiting field-filter $(i,f)$ at $t$, depending on seeing, sky brightness, and airmass. $f_6(i,f,t) = C_m + 2.5 \log (\frac{0.7}{\sigma(i,t)}) + 0.50 (br(i,t)-21) - K(i,f) am(i,t)$ where $C_m$ is a scaling coefficient.}\\ \hline
$f_7(i,t)$  &  \pbox{0.85\textwidth}{hour angle of field $i$ at $t$.}\\ \hline
\hline
\end{tabular}
\end{table}\label{tab_features}

\subsection{Basis functions of the telescope scheduler}\label{sec_lsst_bfs}

Basis functions are fully determined by the value of the features, and are denoted by $\Phi_k(f_1(i,f,t),\dots, f_7(i,f,t))$, for $k = 1 \dots 5$. Each $\Phi_k$, is indexed by a triplet of $(i,f,t)$, time, field, filter, and is evaluated at every decision making step, $t_j$, for all field-filter pairs $(i,f)$. Similar to the update procedure of the features, for a decision at time $t_j$, all six basis functions should be evaluated for all pairs of $(i,f)$, except for the field-filters that are infeasible at $t_j$. Feasibility of a field-filter can be evaluated after the evaluation of the features, by applying the constraints of the region that the field belongs to (see Section~\ref{sec_cstr} for the list of constraints). Thus, while it is required to evaluate the features for all possible pairs of $(i,f)$ at all decision steps, the number of the basis function evaluation is approximately a factor of three less than the number of the features.

\textit{Common basis functions} are shared amongst all of the regions. They are designed to reflect the five general decision factors described in Section \ref{sec_lsst_problem}. The exact definitions of the common basis functions are reflected in Table~\ref{tab_commonBF}, and the key terms and notations can be found in Table \ref{tab_notatopn}. Note that In the definition of $\Phi_1$, scale factor $s_1= 0.43$ is empirically evaluated to ensure that 80\% of the observed values of this basis function, for the visited states, in the simulations are between 0 and 1. Without loss of generality, scaling the values of the basis functions is a regulation that improves the rate of training convergence with a uniform numerical scale of the solution path.

\begin{table}[h]
\caption{Basis functions of the Feature-Based scheduler, the building blocks of the decision function.}
\begin{tabular}{| l | l |}
\hline
Notation & Definition\textbackslash Description\\ \hline \hline
$\Phi_1(f_1(i,f,t))$ & \pbox{0.84\textwidth}{$s_1.f_1(i,f,t)$, the cost of the required time for visiting field-filter $(i,f)$.}\\ \hline
$\Phi_2(f_2(i,f,t))$ &\pbox{0.84\textwidth}{$\begin{cases}0.5,& \text{if } \sum\limits_{f}{f_2(i,f,t)} = 0\\ 1,& \text{if } \sum\limits_{f}{f_2(i,f,t)} = 2\\ 0,  & \text{else,}\end{cases}$, \newline reflects the short term visit/revisit priority of field $i$, conditioned on the total number of the previous same-night visits.}\\ \hline
$\Phi_3(f_5(i,t))$ &  \pbox{0.84\textwidth}{$(1 - \frac{f_5(i,f,t)}{\max_\iota \max_\phi f_4(\iota,\phi,t)})$, reflects the long-term visit priority of field-filter $(i,f)$, based on the ratio of its co-added depth to the maximum co-added depth of all pairs of field-filter until $t$.}\\ \hline
$\Phi_4(f_6(i,f,t))$ &\pbox{0.84\textwidth}{ $1 - Pr(f_6(\iota,\phi,t) \leq f_6(i,f,t))$, empirical complimentary CDF of $5\sigma$-depth of all $(i,f)$ pairs at $t$. $\Phi_4$ assigns a cost to field-filter $(i,f)$ based on its relative visiting quality compared to the other field-filter pairs at $t$.}\\ \hline
$\Phi_5(f_6(i,f,t))$ &  \pbox{0.84\textwidth}{$\frac{|hr(i,t)|}{12}$, encourages visiting of the fields near the meridian.}\\ \hline
\end{tabular}
\end{table}\label{tab_commonBF}

As mentioned in Section~\ref{sec_lsst_problem}, the LSST's mission poses different requirements on different regions of the sky.  First we modify $\Phi_2$, for the Wide Fast Deep (WFD) and North Ecliptic Spur (NES) regions. Because they require the telescope to observe a field twice at a same night, and within a valid time window, $[W_1,W_2]$. The following modification prioritizes the fields that have received a first visit, but not a second visit.

\begin{equation*}
\Phi_2^{\text{WFD}}(f_2,f_3,f_4) =\begin{cases} \Phi^{\text{pair}}(f_3)I_{f_4> W_2},& \text{if } \sum\limits_{f}{f_2} = 1,\\ \Phi_2(f_2),& \text{else,} \end{cases}
\end{equation*}
where, $\Phi^{\text{pair}}(f_3) = \exp(- \frac{\min_{\phi}f_3(i,\phi,t)}{W_2})$, if $ \min_{\phi}f_3(i,\phi,t) < \frac{W_2}{2}$, to rank the fields that have received their first visit of the night (hence the condition). For any other cases, $\Phi^{\text{pair}}(f_3)$ is zero.

For the Deep Drilling Field (DDF) survey which contains a very small fraction of the visible sky's area, instead of adjusting the basis functions, we treat the DDFs as interruptions to the scheduler operations (with each interruption comprising a sequence of DDF observations). Fortunately, the recoverability attribute of the Feature-Based scheduler enables the scheme of the interruptions to be a part of the decision making procedure as long as the interruptions are not too frequent.

\subsubsection{Controllability of the scheduler}\label{sec_sim_cont}

As discussed in Section~\ref{sec_SM}, the mission's objective has to be controllable via the design parameters, $\theta$. In this section we present an empirical observation of the mission's objective controllability. If the value of the objective function does not sufficiently respond to the variations of the design parameters, it is a sign of poor choice of the basis functions and/or input information. Either of these factors lead to a structure that does not admit a sufficiently optimal solution, and if on the other hand, the objective function is extremely variable with respect to the changes in the design parameters, the solution of the training is not reliable because the objective is not a well-behaved function of the optimization variables.

\begin{figure}[h]
\begin{center}$
\begin{array}{ll}
\includegraphics[width=.5\linewidth]{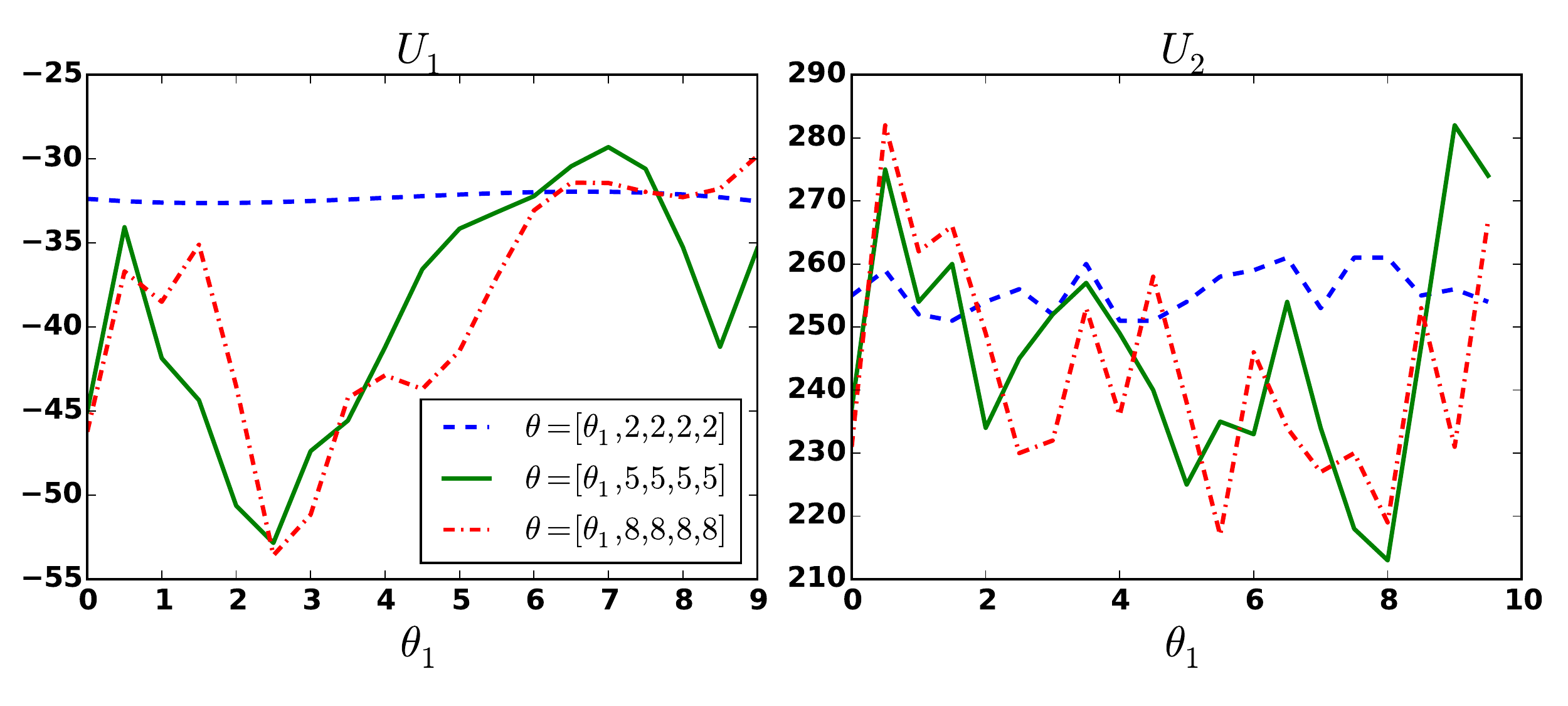}&
\includegraphics[width=.5\linewidth]{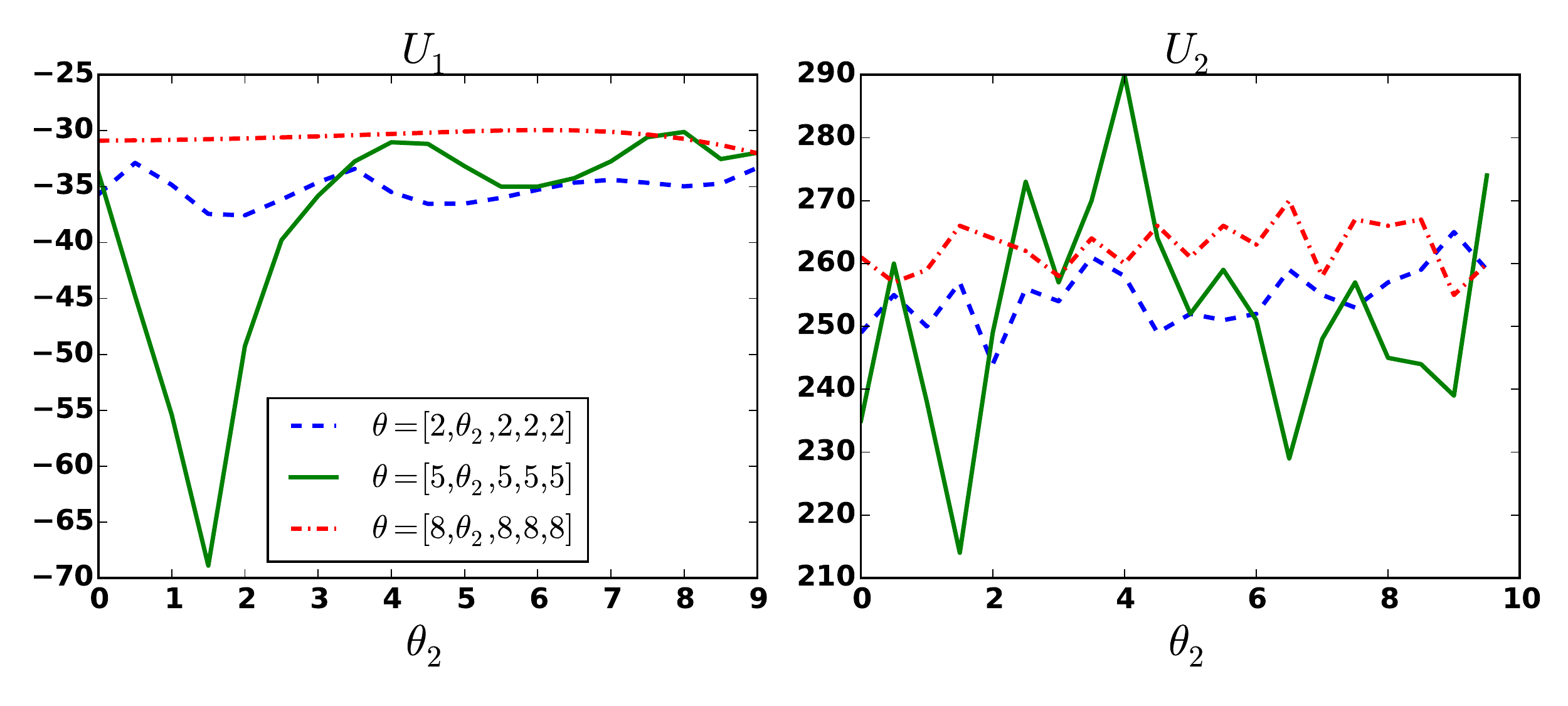}
\end{array}$
\end{center}

\begin{center}$
\begin{array}{ll}
\includegraphics[width=.5\linewidth]{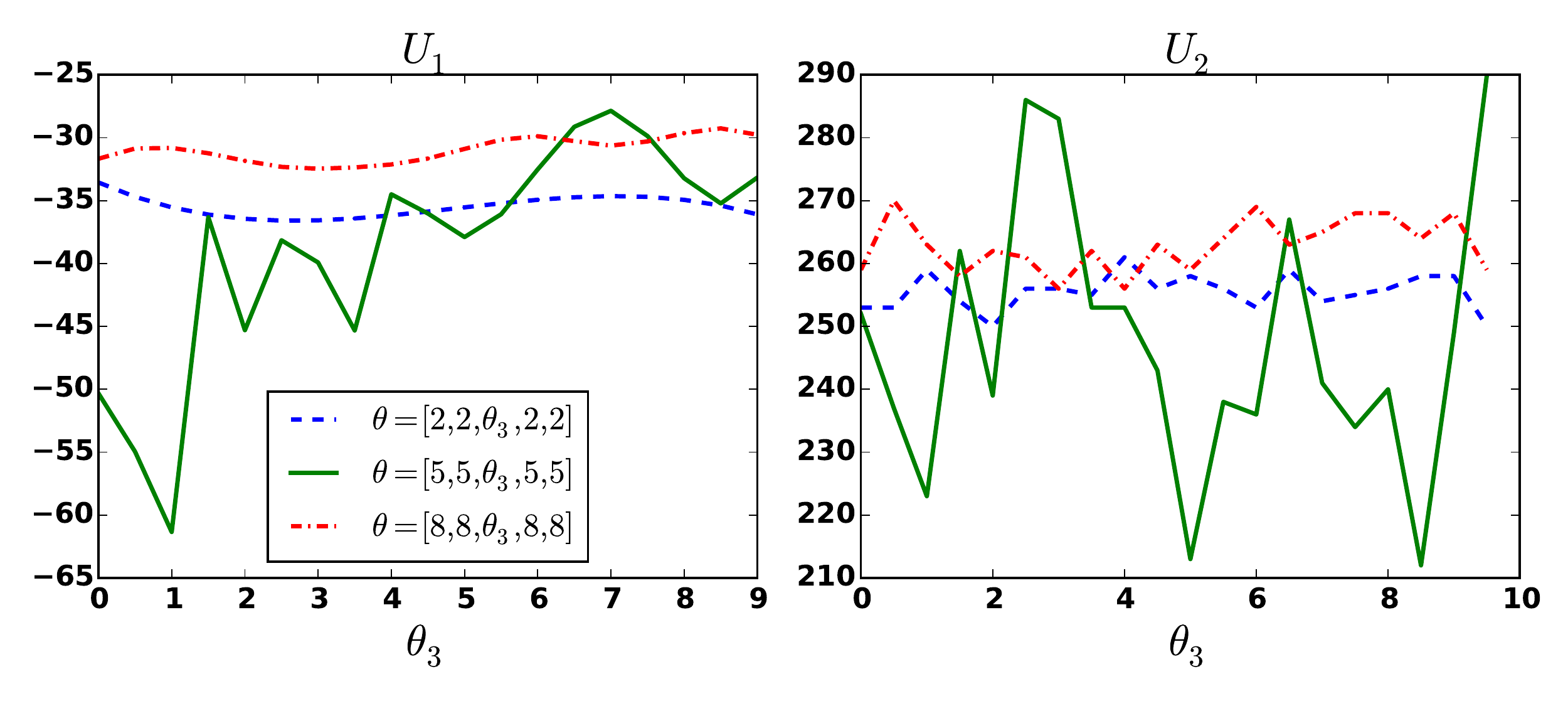}&
\includegraphics[width=.5\linewidth]{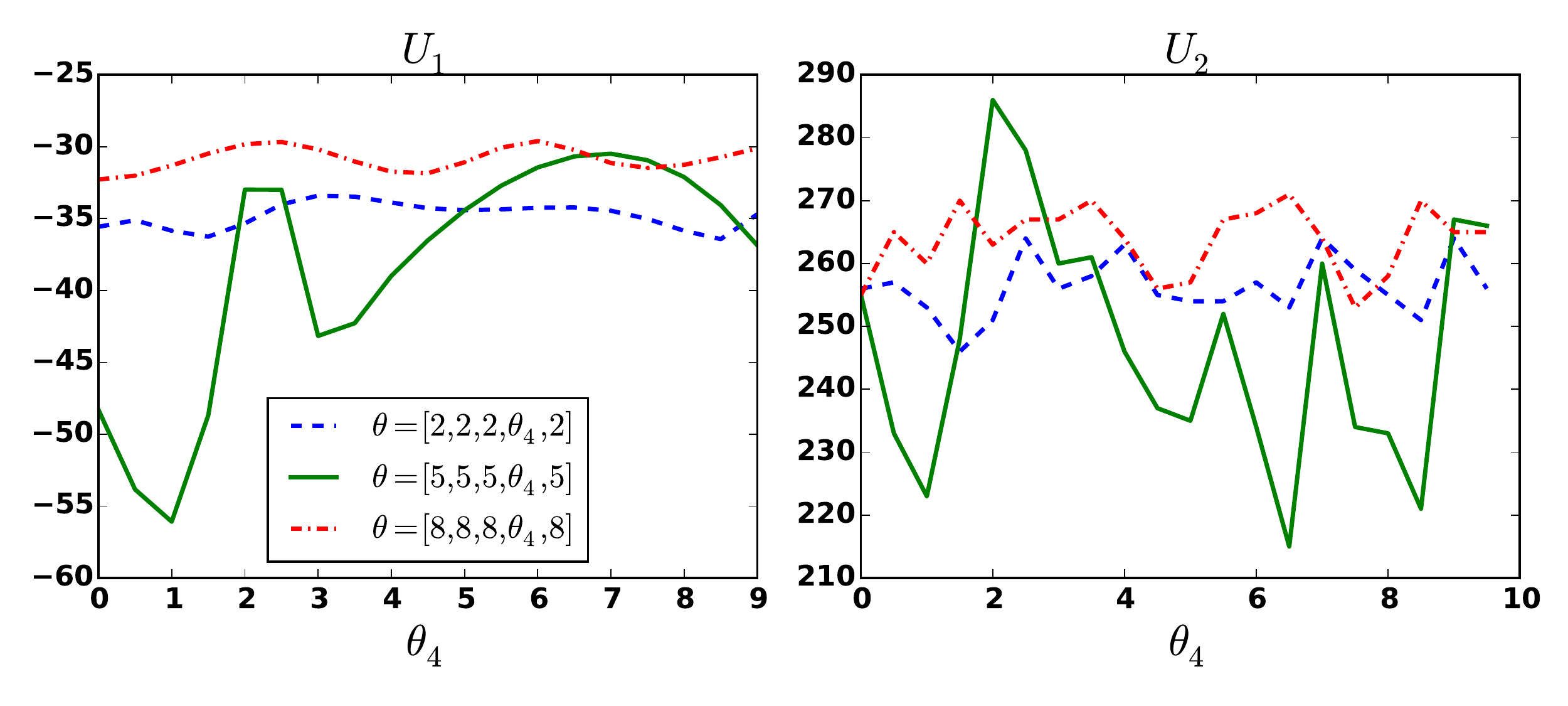}
\end{array}$
\end{center}

\begin{center}$
\begin{array}{l}
\includegraphics[width=.5\linewidth]{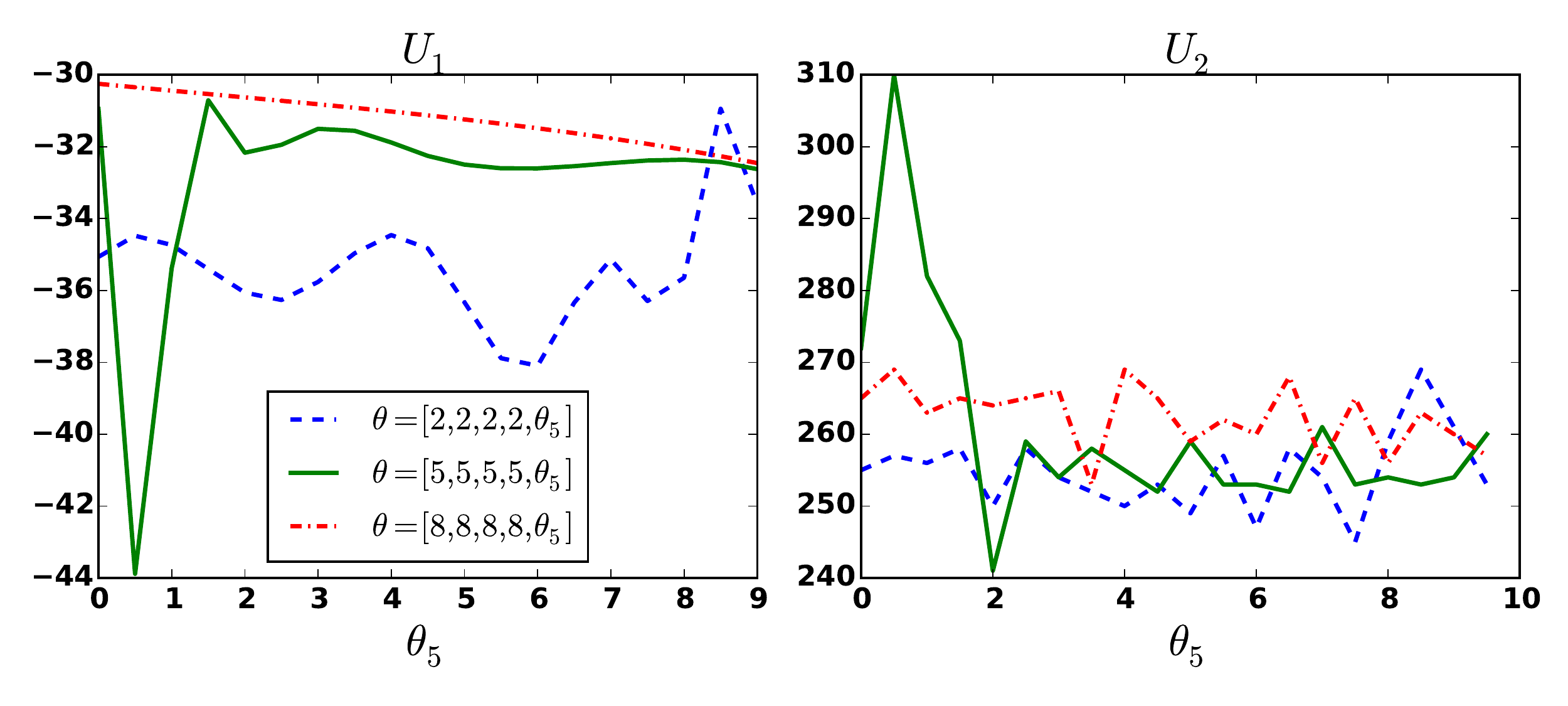}
\end{array}$
\end{center}

\caption{One dimensional slices of the two simple objective functions (\ref{equ_short_term_U1}) and (\ref{equ_short_term_U2}). The variation of the objective functions, specially in the mid-range slices (solid line) suggests that the scheduler's performance is controllable with the design parameters $(\theta_1,\theta_2,\theta_3,\theta_4,\theta_5)$ within the proposed range of the search. This is of course valid for the performance that is measured with either of $U_1$, or $U2$.}
\label{fig_controlability}
\end{figure}

Fig~\ref{fig_controlability}, shows a few one dimensional slices of the two following simple objective functions (\ref{equ_short_term_U1}) and (\ref{equ_short_term_U2}), evaluated after about five hours of scheduling simulation, with different $\theta$'s in $[0,10]$ range ($\theta$'s are the scheduler parameters and determine its behavior). To observe the variability of the objective function with changes in the $i$'th basis function, we define a sequence of equidistant values for $\theta_i \in \{\theta_i^{j_1},\dots,\theta_i^{j_m}\}$ and keep the other $\theta$'s fixed, then run the scheduler for $T = 4.8$ hrs with all resulting $\theta$'s that are different by the $i$'th element, then evaluate $U_1$, and $U_2$, defined as follows. $U_1$ reflects the slew time, $slew()$, and airmass, $am()$, aversion, and $U_2$ reflects the time-efficiency of the operation by counting the total number of observation.

\begin{equation}\label{equ_short_term_U1}
U_1(x_0,x_{1}, \dots, x_{n})= -\sum_{\{i:t_0<t^i<t_n\}} {slew(id(t_{i}), id(t_i)) + 10 am(id(t_i))}
\end{equation}

\begin{equation}\label{equ_short_term_U2}
U_2(x_0,x_{1}, \dots, x_{n})= n
\end{equation}

Figure~\ref{fig_controlability}, contains slices of the 5-dimensional $U_1$ and $U_2$. Both of the simple objective functions reasonably respond to the changes in all five dimensions of the variable $\theta$, which can be an evidence of the controllability of the objective function. Moreover, the smaller variations for slices closer to the boundaries of the search space, suggest that the design and scaling of the basis functions provide a desirable behavior within the proposed search space.

\subsection{Survey-specific constraints}\label{sec_cstr}

The scheduler's decision at each time step is an admissible (feasible and measurable) pair of field-filter $(i,f)$, thus before each decision, one needs to specify the set of feasible actions. Feasibility of a candidate pair is driven by the following measurable factors:

\begin{itemize}
\item Visibility: The candidate field-filter has to be visible.
\item Quality: The expected observational quality of a field-filter has to be better than the given lower threshold.
\item Survey's timing: The science-driven revisit constraints has to be respected.
\end{itemize}

Exact expression of the proposed constraints for the LSST scheduler are presented in Table~\ref{tab_feasibility}.

\begin{table}
\caption{Feasibility of field-filter $(i,f)$ for a visit at $t^{n+1}$, as evaluated at $t^n$.}
\begin{tabular}{| l | l | l | l |}
\hline
& Constraints& Description & region\\ \hline \hline

1& \pbox{0.3\textwidth}{$\tau_{rise}(i,f,t^{n+1}) \leq t^{n+1} \leq \tau_{set}(i,f,t^{n+1}) $ }& \pbox{0.5\textwidth}{field-filter $(i,f)$ has to be above the acceptable airmass horizon at $t^{n+1}.$ }& All regions\\ \hline

2& \pbox{0.3\textwidth}{$ E[f_4(i,f,t^{n+1})] \neq 0 $ }& \pbox{0.5\textwidth}{field-filter $(i,f)$ is not temporarily masked (e.g. by the moon) at $t^{n+1}$. }& All regions\\ \hline

3 &  \pbox{0.3\textwidth}{$\sum_{f}f_2(i,f,t^n) < N^{\textit{survey}}$ }& \pbox{0.5\textwidth}{$N^{\textit{survey}}$ poses a region dependent upper-bound on the number of the visits for each field. $N^{WFD}= N^{NES} = 3$, and $N^{GPR}= N^{SCP} = 1.$ }& All regions\\ \hline

4 &  \pbox{0.3\textwidth}{$E[f_6(i,f,t^{n+1})] < \sigma({\textit{survey},f})$ }& \pbox{0.5\textwidth}{the expected quality of visiting field-filter $(i,f)$ at $t^{n+1}$ has to be better than the given threshold, $\sigma(.)$, that depends on the survey and the filter. }& All regions\\ \hline

5&  \pbox{0.3\textwidth}{$f \neq id(t^n)$ }& \pbox{0.5\textwidth}{consecutive visit of a same field is not allowed. }& All regions\\ \hline

6&  \pbox{0.3\textwidth}{if $\sum_{f}f_2(i,f,t^n) = 0$ then \newline $\max_\phi f_4(i,\phi,t^n) > \frac{W_1+W_2}{2}$ }& \pbox{0.5\textwidth}{the first visit of field $f$ has to occur $\frac{W_1+W_2}{2}$ time before it becomes invisible, so that the second visit of $f$ can be scheduled in the valid time window. }& WFD and NES\\ \hline

7&  \pbox{0.3\textwidth}{if $\max_{\phi}\theta(i,\phi,t^n) > \tau_s(t^n)$ then \newline $ W_1 \leq \min_{\phi}f_2(i,\phi,t^n) \leq W_2 $}& \pbox{0.5\textwidth}{if there has been a same-night visit of field $f$ until $t^n$, then the next same-night visit has to occur in the valid time window. }& WFD and NES\\ \hline

8&  \pbox{0.3\textwidth}{if \newline $\max_{\phi}\theta(i,\phi,t^n) > \tau_s(t^n)$ then \newline $f \notin \{\text{y},\text{u}\}$}& \pbox{0.5\textwidth}{if there is a same-night visit of field $f$ until $t^n$, then the next same-night visit cannot be with either of u or y filters. }& WFD\\ \hline

9&  \pbox{0.3\textwidth}{$f \notin \{\text{y},\text{u}\}$ }&  \pbox{0.5\textwidth}{visits with u filter and y filter is not allowed. }& NES\\ \hline

\end{tabular}
\end{table}\label{tab_feasibility}

\subsection{Scheduler optimization}\label{sec_lsst_opt}

In this Section, we present two simple choices of high-level mission objectives to demonstrate the application of the proposed optimization approaches, discussed in Section~\ref{sec_opt}. (More sophisticated mission objective functions can be defined based on LSST performance studies such as \citep{2016AJ.151..172G}, \citep{2018AJ.155..1G}, \citep{2017AJ.153..186J}, and \citep{2012AJ.144..9O}). The choice of optimization algorithm depends on the nature of the mission objective. The first experimental mission objective function in this section can be expressed as the discounted sum of instant rewards $R(s_{i-1}, s_i)$, thus the reinforcement learning is applied to find the scheduler's parameters $\theta$. The second objective function cannot be decomposed as a discounted sum of instant rewards, thus we used the global optimizer approach. From the computational point of view, the first approach is preferred. For the following experiment, the reinforcement learning is about 10 times faster than the global optimization, and requires 50 times less memory. From the practical point of view however, for some missions, it is impossible to define an objective function that can be expressed as discounted sum of instant rewards. In which case, the mission objective can be optimized only via global optimizer.

In the following experiments, for both of the optimizations we used a simulated model of the telescope \citep{2014SPIE.9150E..14C} and the environment, including the brightness of the sky and coverage of the clouds which are developed based on the measurements at the LSST site. 

\subsubsection{Reinforcement learning for the first choice of mission objective.} 

Let the instant reward, $R(i-1,i)$, be $-slew(id(t_{i-1}), id(t_{i})) - am(id(t_i))$. It is defined to be a linear combination of the slew time to point the telescope from the $(i-1)$th field to the $i$th field, and the airmass of the destination. Since both factors have negative effect on the quality of the observation, the reward would be measured by the negative of each. Then the mission objective function can be simply defined as  $\sum_{i=0}^N \gamma^i R(i-1, i)$.

\begin{figure}[h!]
\begin{center}
\includegraphics[width=0.5\linewidth]{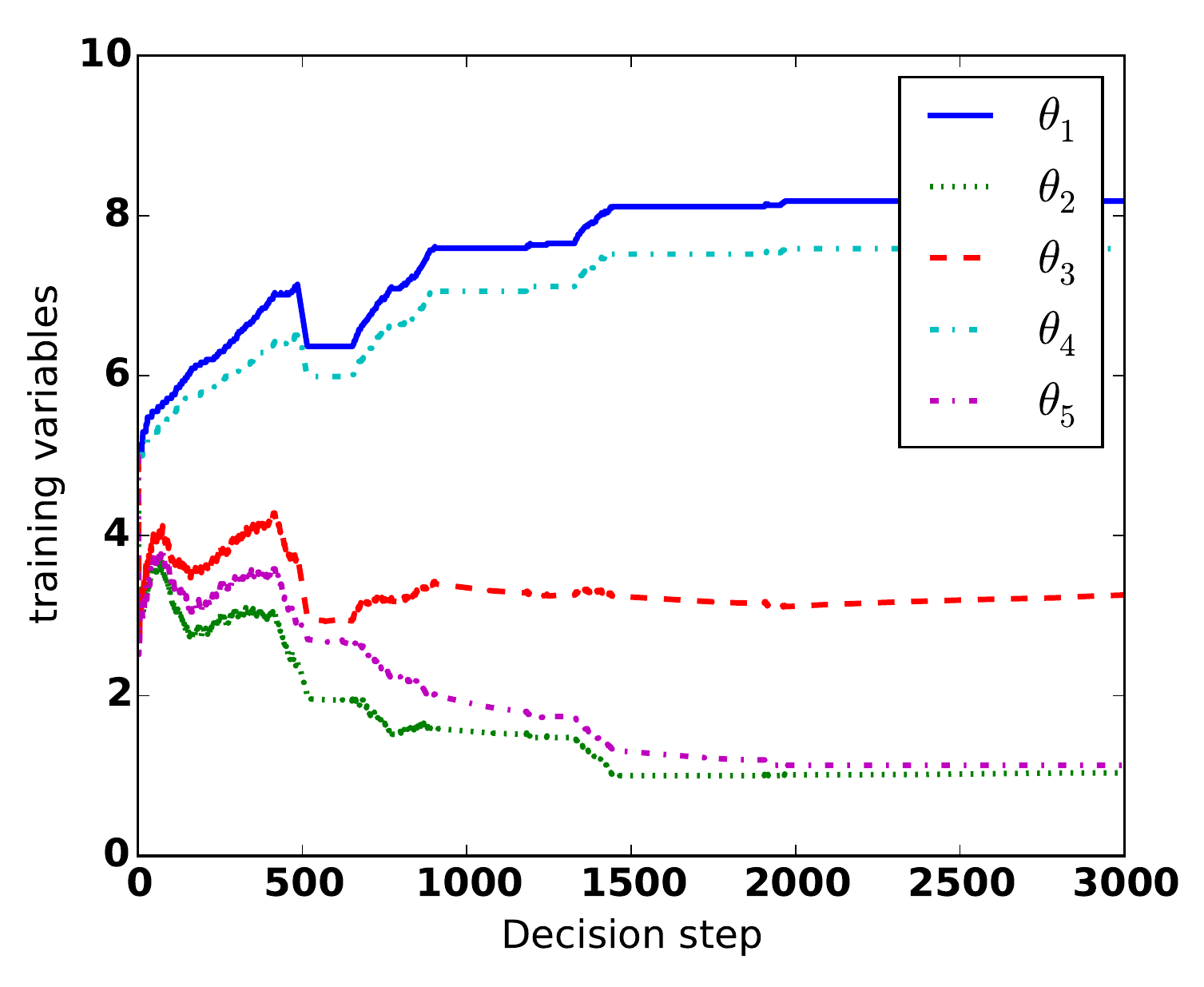}
\end{center}
\caption{Reinforcement learning of the scheduler's parameters, $(\theta_1,\theta_2,\theta_3,\theta_4,\theta_5)$. All of the parameters are initialized at the mid-range value, 5. During the simulation, at each step there is a reward associated with the decision which implies a small adjustment on each of the five parameters. Then the next decision will be taken with a slightly different set of variables. This procedure continues until the adjustments on the variables are negligible.}
\label{fig_theta_conv}
\end{figure}

The simulation for the reinforcement learning starts at $t_0 = 2462867.5~mjd$ (2021 January 1), with $\theta^0 = (5,5,5,5,5)$, initialized at the mid-range values, and continues until $\theta$ converges. Figure~\ref{fig_theta_conv}, is the training curve for all of the variables over a course of 3000 decisions, . The discount rate $\gamma = 0.9$, and learning rate $\frac{0.01}{\log^3(i)}$ are chosen empirically. 

For the above choices of reward, learning rate, discount factor and initialization, $\theta$ converges to $\theta^* = (8.18,  1.04,  3.26,  7.59,  1.13)$.
In this approach, the computational time to evaluate the update is negligible with respect to the time required to evaluate the features, therefore, the computational time of the reinforcement learning is almost equal to the scheduling simulation (without the updates on $\theta$) which is linear with respect to the number of the decisions. With a personal computer\footnote{Processor: 1.6 GHz, Memory 1600 MHz DDR3}, each decision takes about $0.8$ second, thus the time of the convergence for the simulation, presented in this section is $3000\times 0.8 \text{ sec } = 40 \text{ minutes }$.

\subsubsection{global optimization for the second choice of mission objective.} One of the important simple objective functions that cannot be expressed as the discounted sum of the rewards is the total number of the observations from any $t_i$ to any $t_j$ that can be expressed as a utility function, $U_{\pi_{\theta}}(x_i, x_{i+1}, \dots, x_j) = j - i$.

To find a set of parameters, $\theta$, that optimize the above objective function we applied the global optimization approach, explained in Section \ref{sec_opt}, with the following regulatory constraints.\\
(1) $\theta \geq 0$: Positive coefficients for the basis functions are assumed in the design of the basis functions. Because in the context of the telescope scheduling, it is more natural to create the basis functions to reflect the cost of the operation.
(2) $\theta_1 =\theta_0$: Without loss of generality, we fix the value of the first element of $\theta$ to reduce the dimension of the optimization problem by one. Because, homogeneity of the policy implies that if $\theta^*$ yields an optimal scheduler, then $\alpha \theta^*$ for $\alpha > 0$ yields an optimal scheduler too.

\begin{figure}[h!]
\begin{center}
\includegraphics[width=0.5\linewidth]{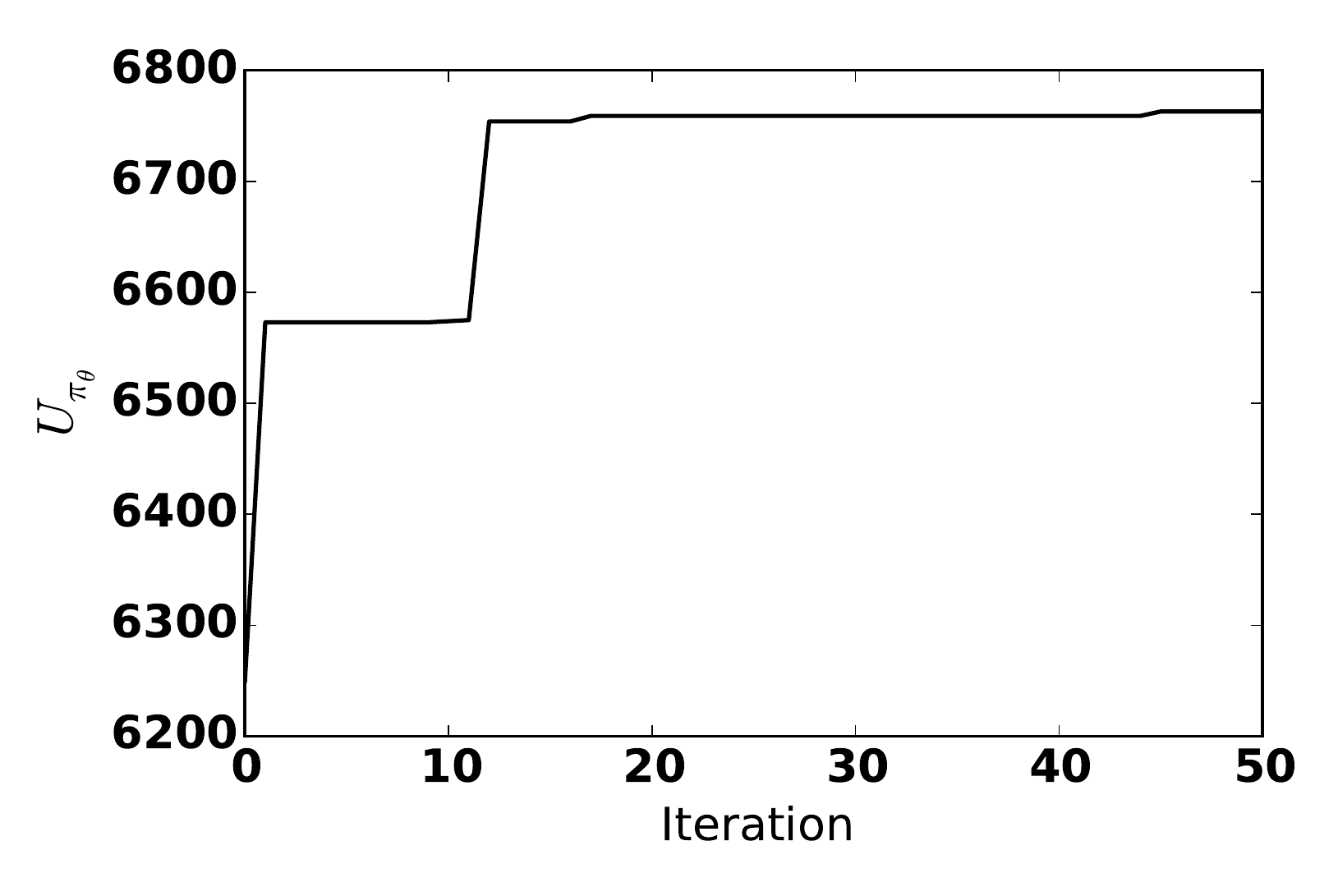}
\caption{Progress of the black-box objective function $U_{\pi_{\theta}}$ over the iterations of the $e$DE algorithm. $U_{\pi_{\theta}}$, for this simulation is the total number of the observations for 10 nights starting from (2021 January 1).}
\label{fig_eDEObjectiveFunction}
\end{center}
\end{figure}

We used the above objective function, $U_{\pi_{\theta}}$, for $t_i = 2462867.5~mjd$ (2021 January 1) and $t_j = 2462877.5~mjd$ (2021 January 11). Figure~\ref{fig_eDEObjectiveFunction}, shows the value of this objective function over the iterations of the $e$DE algorithm. The solution $\theta^* = (1.00, 0.84, 0.99,  1.34,  3.04)$, yields the best $U_{\pi_{\theta}}$ after 50 iterations for $\theta_1 = 1.00$.

$e$DE is a population-based metaheuristic algorithm, and for the result shown in Figure~\ref{fig_eDEObjectiveFunction}, the number of population $N_P$ is set to be 50. Each function evaluation is in fact, the simulation of 10 days of scheduling with a candidate scheduler which takes about 8 minutes, therefore each iteration, in total, takes $N_P * 8$ minutes with a personal computer. The optimization can be manually terminated if the result is satisfactory, or can be continued until a full convergence is achieved. In $e$DE (and all genetic algorithms in general), function evaluation for each individual is independent from other individuals, therefore the parallel implementation of the same algorithm can be faster up to a factor of $N_p$.

\section{Performance of a Modified Feature-Based scheduler for LSST}\label{sec_comp}
In this section, the LSST Metric Analysis Framework \citep{jones2014lsst} is used, to compare\footnote{The sky background models and weather downtime used to benchmark the algorithms are not exactly identical because of the practical difficulties in the separation of the environment and the baseline scheduler implementation at the time. However, for the purpose of the comparisons in this paper, the behavior of our sky and observatory model is sufficiently close to the official model. See \citep{2016SPIE.9910E..13D}, and \citep{2016SPIE.9911E..25R} for the official operations simulator.} the performance of a modified version of the Feature-Based scheduler with opsim V4 and opsim V3, the most recent baseline schedules of the LSST, over a 10-year period of scheduling simulations.\\
The \textit{Modified Feature-Based} scheduler is under active development\footnote{GitHub repository: \url{https://github.com/lsst/sims_featureScheduler}.}, and addresses the observational details of the LSST's mission through the adjustment of the constraints and the basis functions. It is designed to produce a software that can be used in practice. 

The default sky tessellation adopted in the baseline scheduler results in 23\% of the sky being covered by more than one field. In Modified Feature-Based scheduler, we adopt a finer discretization of the sky, and do not require the partitions to be necessarily sized based on the telescope's field of view. The fact that the policy is not computationally expensive to evaluate makes it possible to use a finer discretization of the feature-space. This approach allows the scheduler to handle the field's overlaps which cause inhomogeneity in the coverage of the sky. 

In addition to adopt a finer discretization, we use a spatial dithering scheme to randomize the final pointing of the telescope by a small amount around the center of the partitions to further assist the homogeneity of the coverage. Adopting the dithering scheme, the median number of observations at a typical point in the sky increases by $\sim15$\%. Dithering is also essential for removing systematic effects for science cases such as measuring galaxy counts, (see \citep{Awan2016} for more details). Moreover, the Modified Feature-Based scheduler uses a separate process to track if an observation will need to be observed in a pair, and a separate processes to decide if a Deep Drilling sequence should be executed by interrupting the normal operation of the telescope.

\subsection{Sky coverage uniformity}
\begin{figure}[h!]
\plottwo{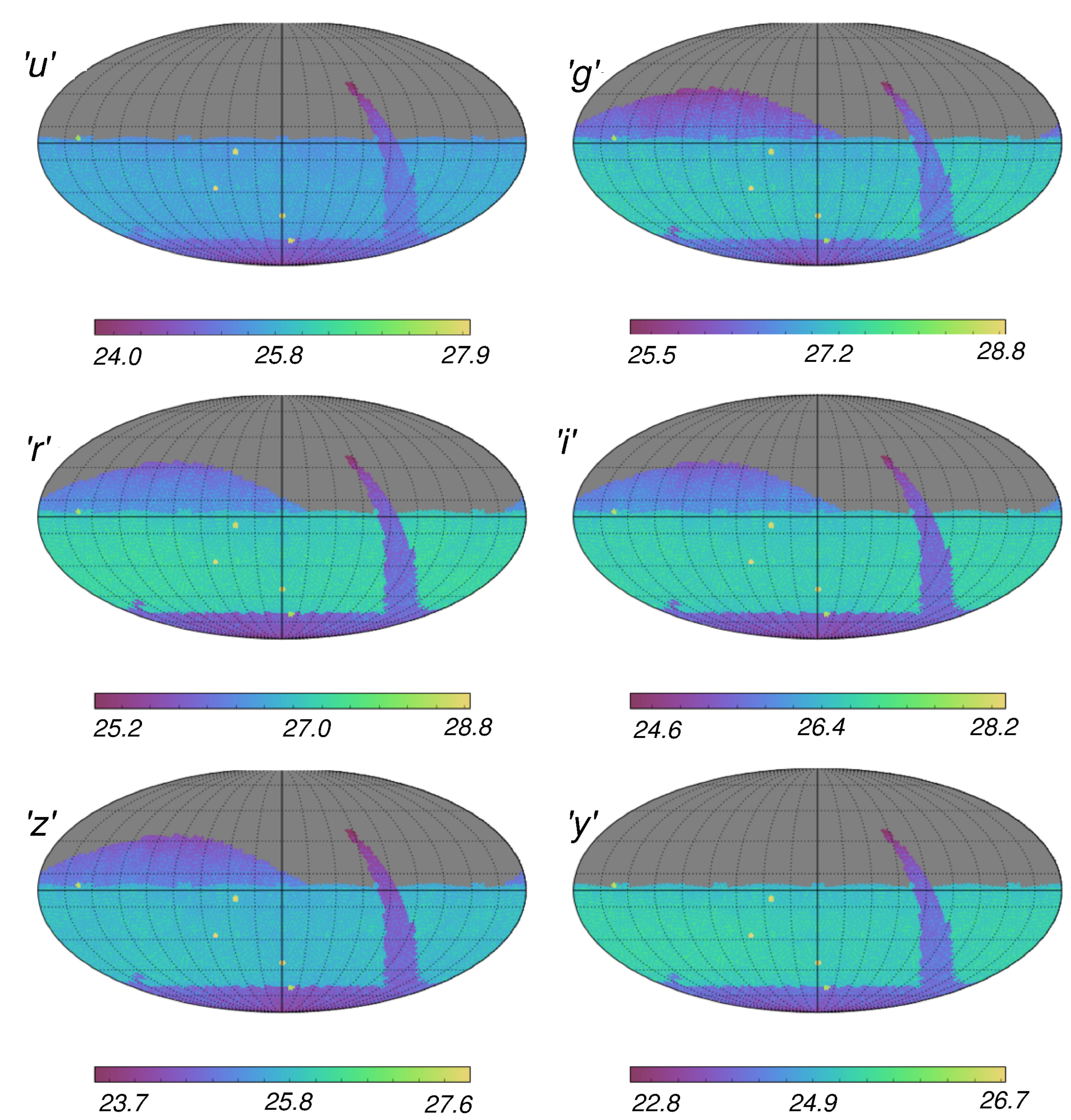}{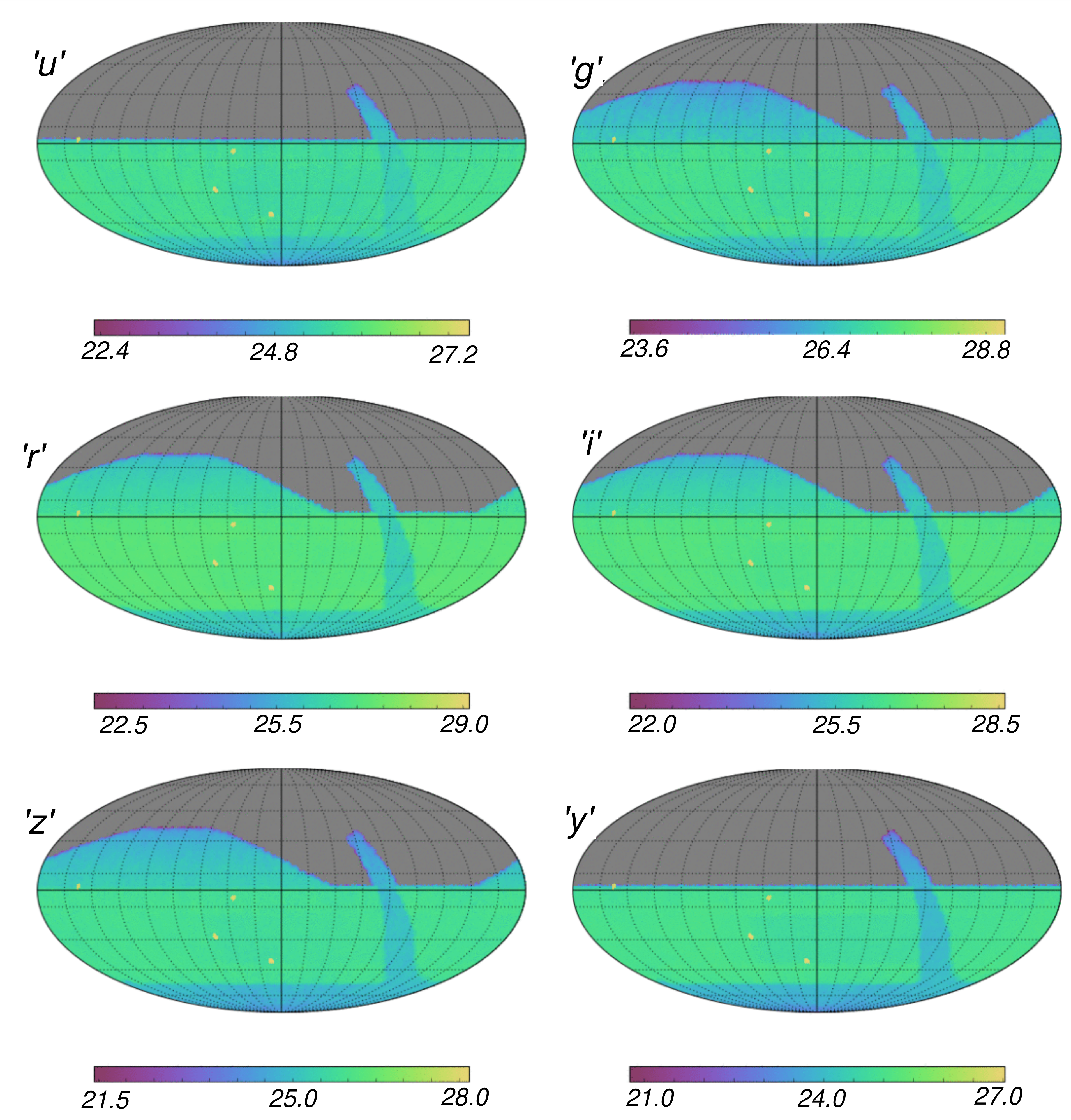}
\caption{The final co-added depth coverage in each of the six filters $\{ u, g, r, i, z, y\}$. According to the mission's objective, the scheduler has to provide a uniform coverage of the visible sky within each region, and in each filter. The left panels show the opsim V4 simulation results while the Modified Feature-Based scheduler is on the right. Even without a given observation proposal, Modified Feature-Based scheduler can closely match the large-scale footprint of the official survey.}
\label{fig_10yrs_skymap}
\end{figure}

\begin{figure}
\epsscale{0.35}
\plotone{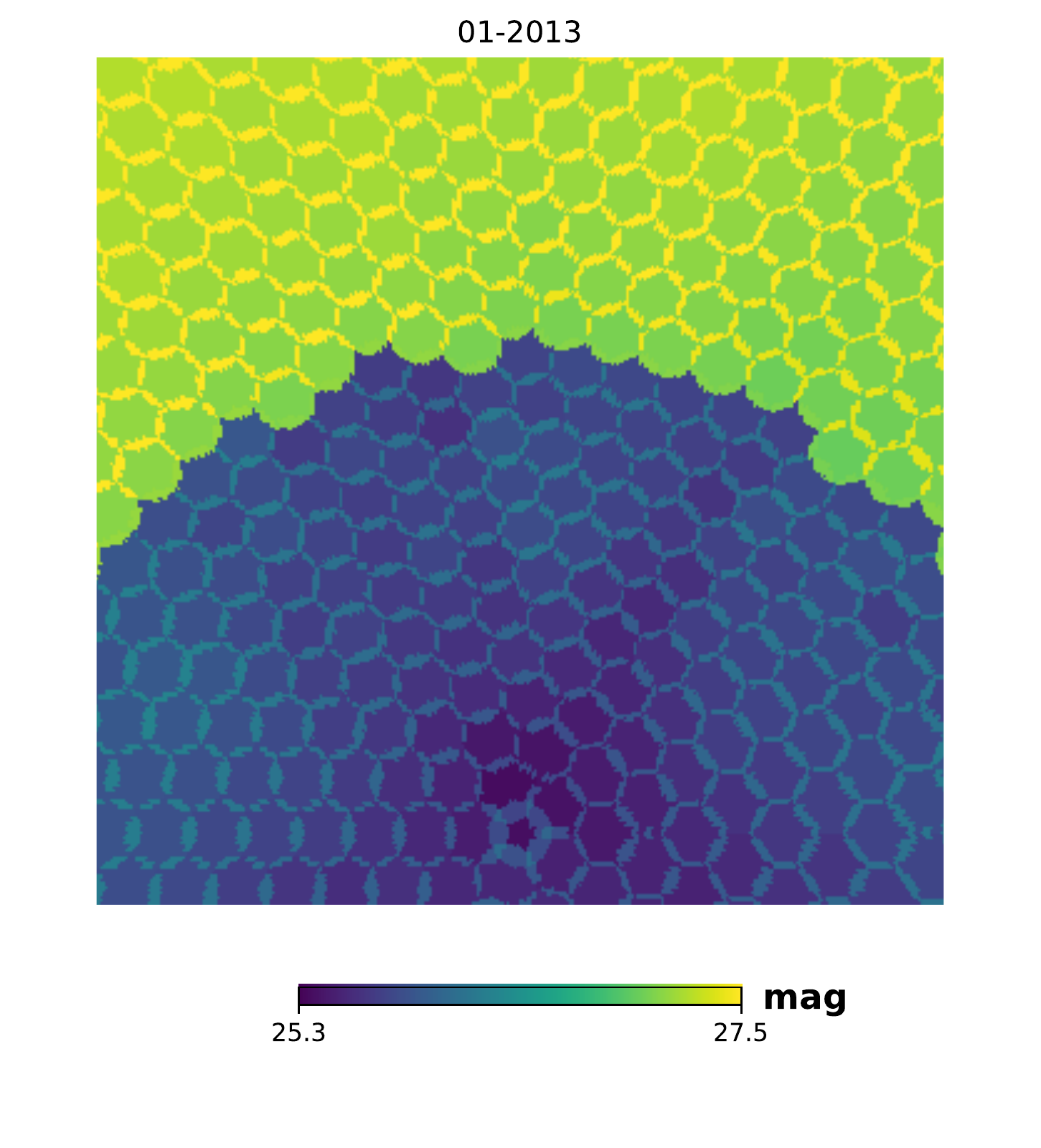}
\plotone{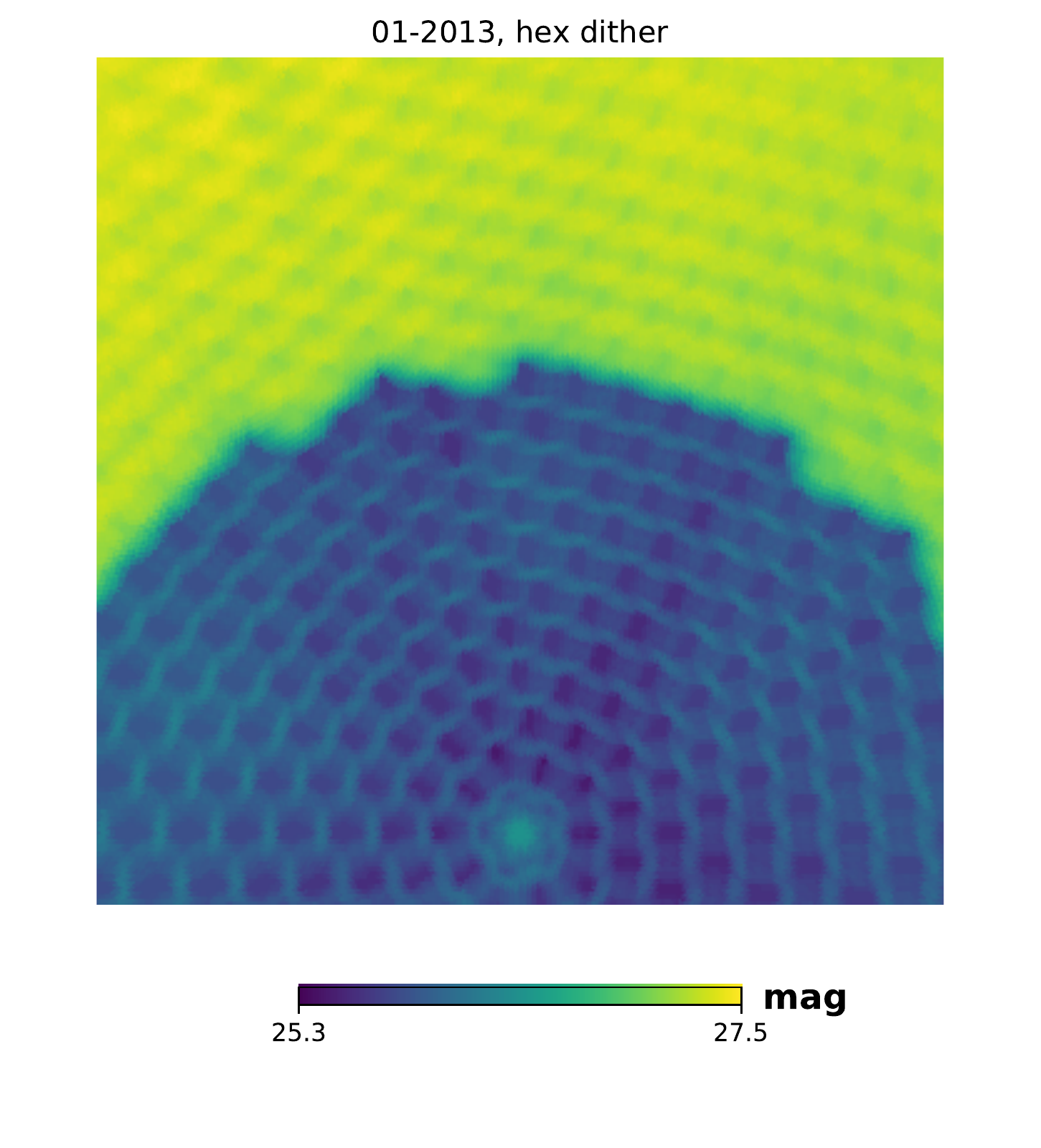}
\plotone{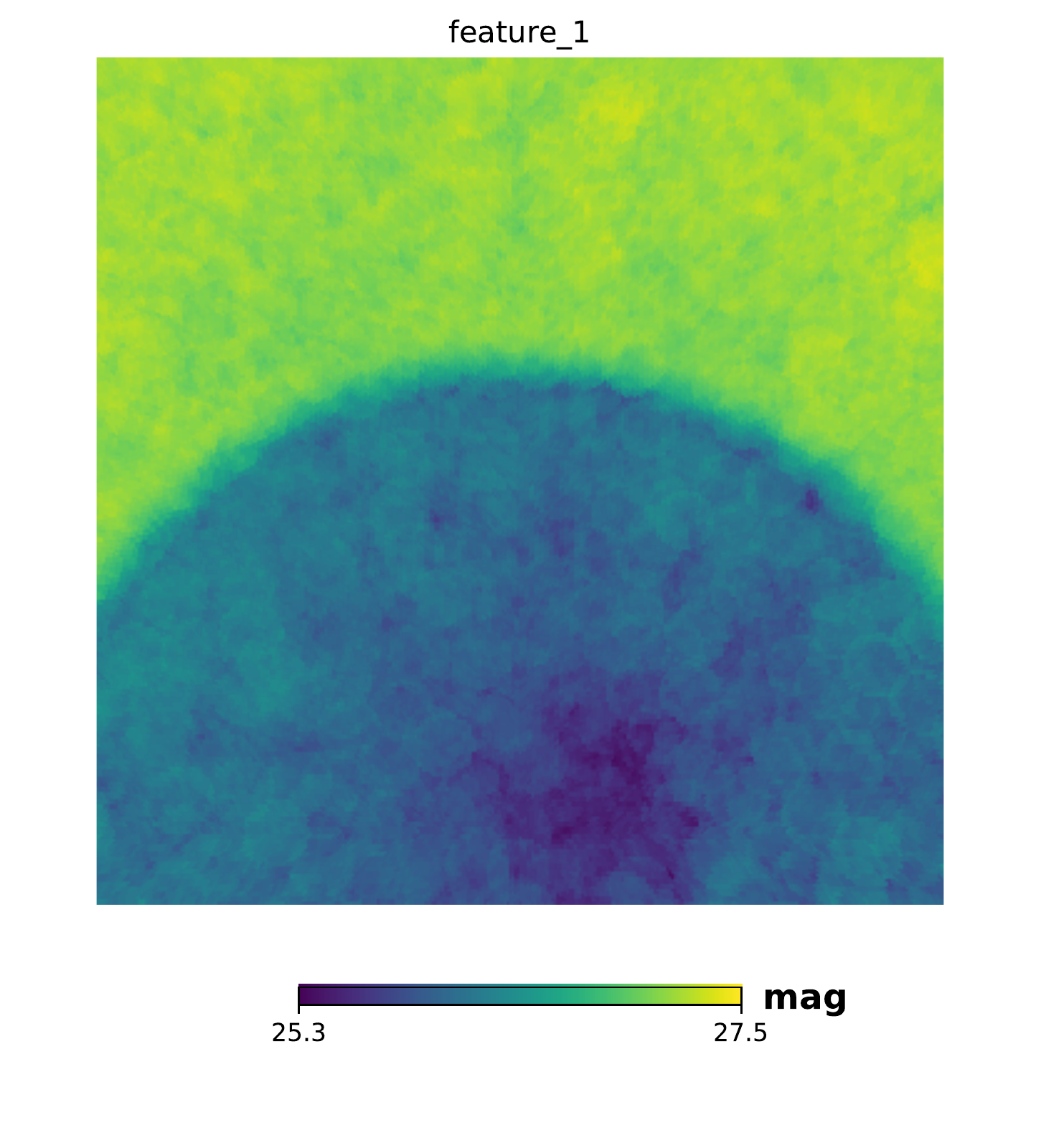}
\epsscale{1}
\caption{Co-added depth coverage in a smaller-scale view of the sky around the border between the WFD (green) and SCP (blue) regions, in the \textit{r} band. The smooth coverage of the Modified Feature-Based scheduler (right) versus the granular pattern of opsim V4 (left), further satisfies the uniformity of the coverage which is one of the most fundamental objectives of the LSST mission. The middle figure is the coverage of opsim V4, with dithering of the same sequence which fundamentally can not become as smooth as the right figure, because unlike Feature-Based scheduler, the scheme of opsim V4 does not easily allow for decision making with arbitrarily fine discretization of the sky.}\label{fig_zoomin_r}
\end{figure}

For a survey telescope, such as the LSST, the density of the co-added depth over the visible sky should ideally be uniform in each filter and within each of the five survey regions. Figure~\ref{fig_10yrs_skymap} compares the values of the co-added depth on a discretized sky map. Figure \ref{fig_zoomin_r}, demonstrates the smoothness of the coverage in a smaller scale for opsim V4, with and without dithering, and compares it with that of the Modified Feature-Based scheduler in the \textit{r} band, all around the boundary of WFD and GP regions. Smoother coverage that the Modified Feature-Based scheduler offers is due to the fine discretization of the sky in the decision making stage, in addition to dithering which is applied after the decision is made.

\begin{figure}[h!]
\centering
\includegraphics[width=1.0\linewidth]{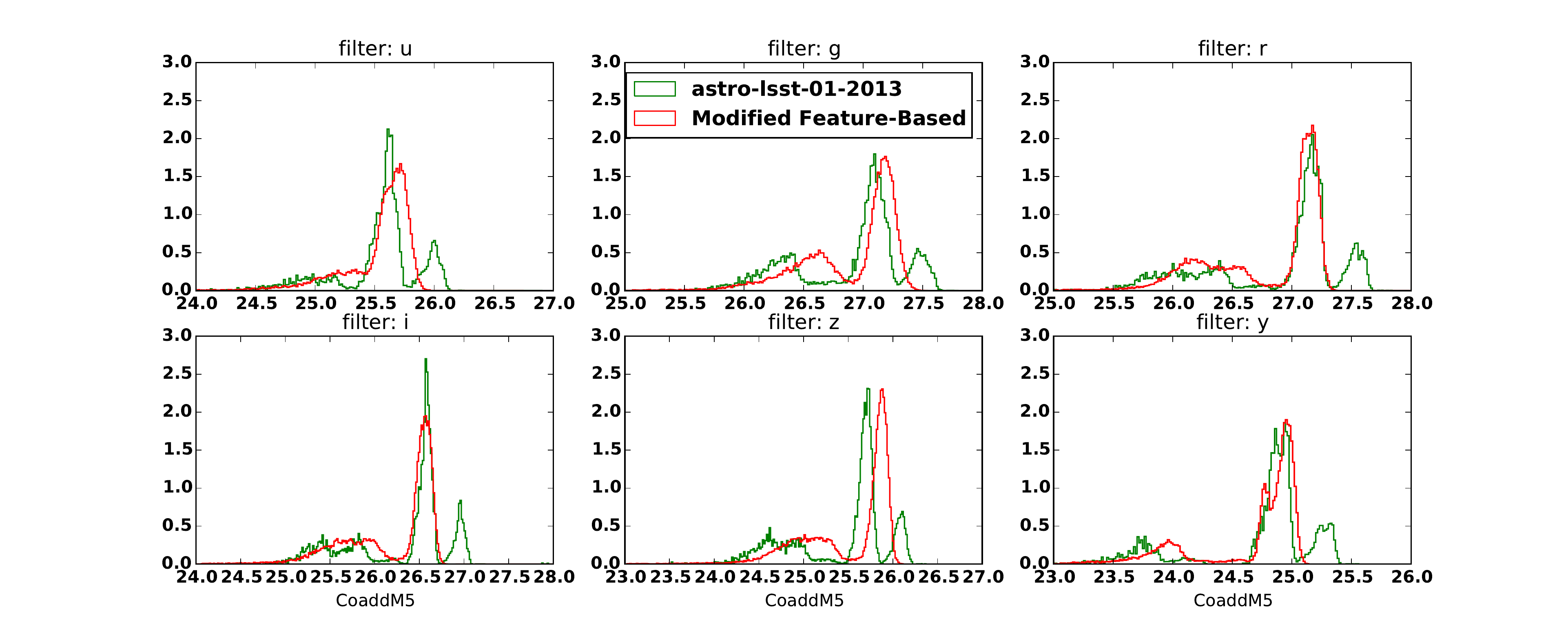}
\caption{Each plot compares the distributions of the co-added depth coverage in one of the six filters. A dithering scheme in the Modified Feature-Based scheduler in addition to a finer tessellation of the sky smoothens the density of the coverage where the fields overlap.}
\label{fig_10yrs_hist}
\end{figure}

Figure \ref{fig_10yrs_hist} compare the distribution of the co-added depth on a (finely) discretized sky. The Modified feature-Based scheduler has paved the left-most peak that appears in the distribution of opsim V4. This peak is the result of the field's overlaps that receive more visits than specified in the configuration of the scheduler. Table \ref{table_10yrs_hist} contains the median and variance of the co-added depth for both schedules in each of the main sky regions and in each filter. Modified Feature-based scheduler provides deeper (higher median), and more uniform (lower variance) coverage in most of the cases.

\begin{table}
\caption{The median and variance of the co-added depth distribution on a finely discretized sky. Modified Feature-Based scheduler closely matches the footprint of the official survey, and in addition outperforms opsim V4 in terms of the uniformity of the coverage, with lower variances, specially in WFD and SCP regions.}\label{table_10yrs_hist}
\begin{center}
\begin{tabular}{l|cccc|cccc} \hline
&\multicolumn{8}{c}{Median, Variance} \\ \hline
&\multicolumn{4}{c|}{opsim V4} &\multicolumn{4}{c}{Modified Feature-Based} \\
filter  & WFD & GP & SCP & NES & WFD & GP & SCP & NES\\
\hline
\textit{u}& 25.63, 0.04 & 25.12, 0.11& 24.91, 0.12& - &25.68, 0.01 &25.32, 0.05 & 25.10, 0.04& -\\
\textit{g}& 27.13, 0.04 & 26.41, 0.09& 26.32, 0.12 & 26.30,0.13& 27.18, 0.01 & 26.69, 0.04&26.56, 0.04 &26.47, 0.09 \\
\textit{r}& 27.19,0.04 & 26.01,0.16& 25.84,0.24 & 26.38, 0.12 & 27.14, 0.01& 26.21, 0.07 & 26.08, 0.05 & 26.43, 0.09\\
\textit{i}& 26.60, 0.04 & 25.44, 0.15 & 25.28, 0.22 & 25.82, 0.12 & 26.56, 0.01 & 25.68, 0.08 & 25.43, 0.07 & 25.88, 0.09\\
\textit{z}& 25.73, 0.04& 24.62,0.17& 24.57, 0.21& 24.90, 0.14 & 25.87, 0.01 & 25.03, 0.09 & 24.81, 0.05 & 25.16, 0.10\\
\textit{y}& 24.92, 0.04 & 23.81, 0.16 & 23.72, 0.21& - & 24.92, 0.02 & 24.01, 0.09 &23.88, 0.06 & -\\
\end{tabular}
\end{center}
\end{table}

\subsection{Pairs}
\begin{figure}[h!]
\plottwo{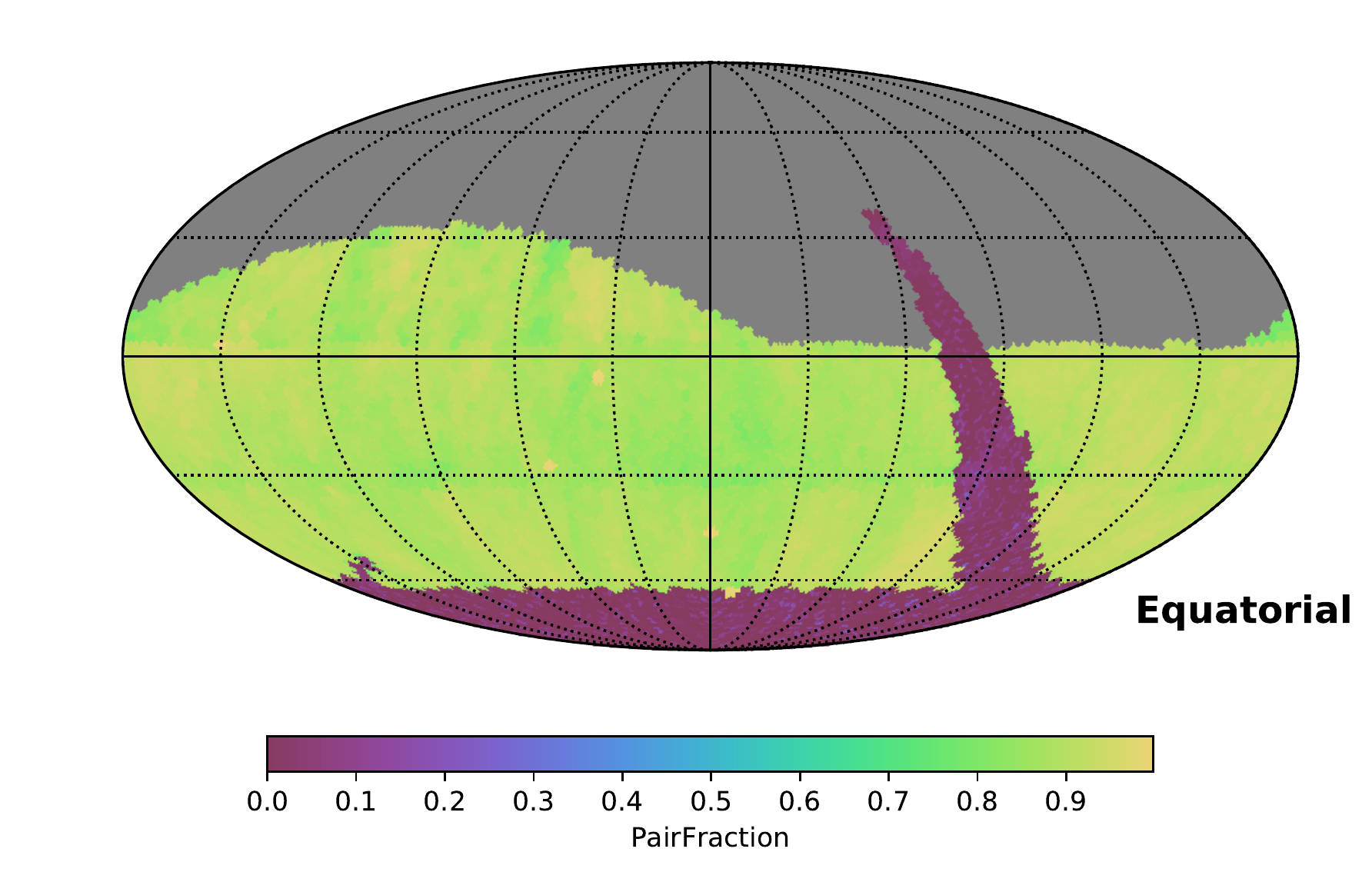}{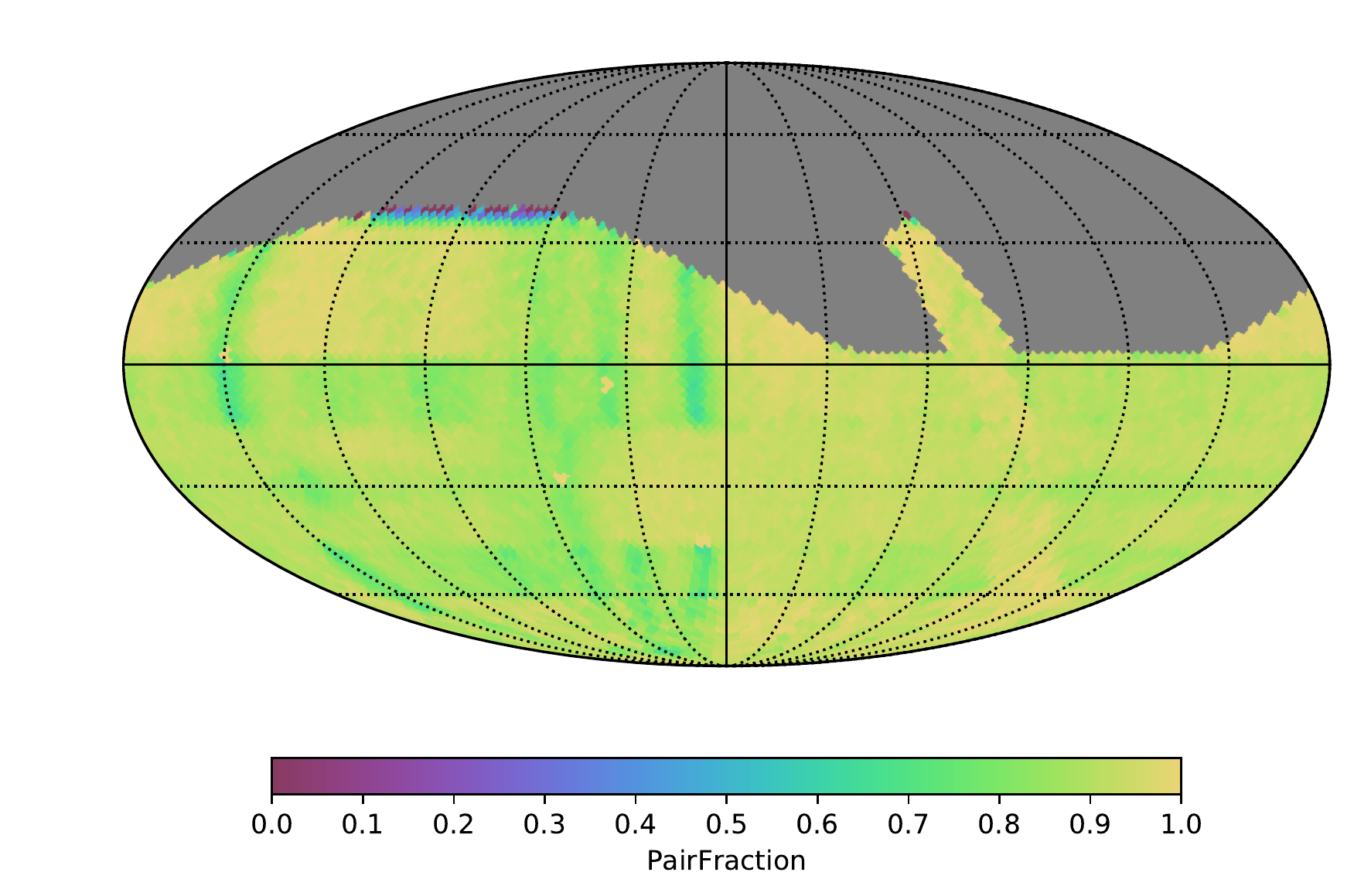}
\caption{The ratio of the pairs (in the $g$, $r$, and $i$ filters) to the total number of the observations on the sky map. For the areas that the pair constraint is applied, the ratio is desired to be one. For Modified Feature-Based scheduler (right), the pair constraint is applied to all of the regions and for opsim V4 (left), it is applied to the WFD and NES regions only.}
\label{fig_10yrs_pair}
\end{figure}

\begin{figure}[h!]
\centering
\includegraphics[width=.4\linewidth]{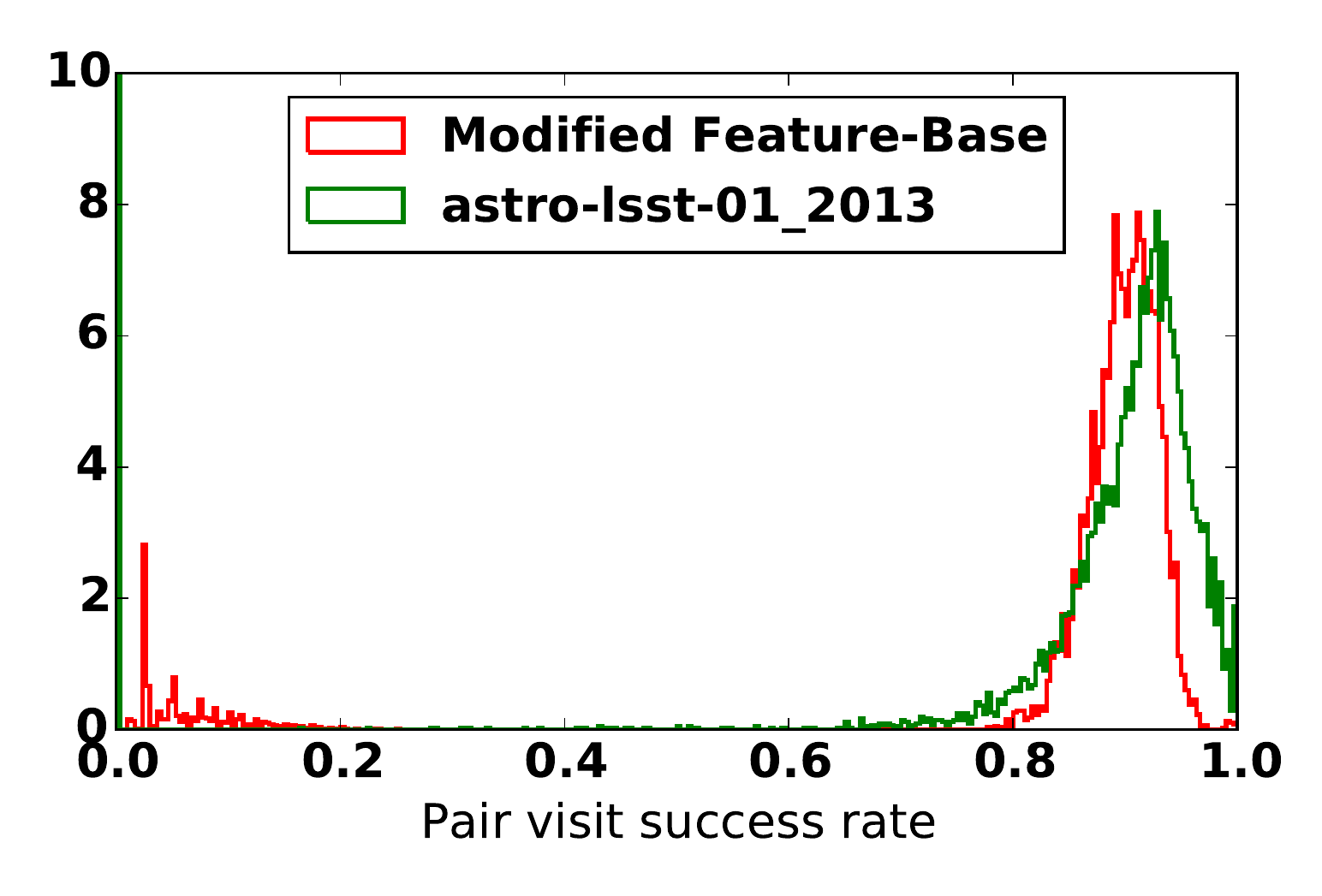}
\caption{Distribution of the pairs ratio to the total number of the observations, in any of the \textit{g}, \textit{r}, and \textit{i} filters. The Feature-Based scheduler in compare with opsim V4, sacrifices perfect pairs observation for a limited area of the visible sky (with a lower mean) to maintain a uniform ratio of pairs for a larger area of the sky (with a sharper distribution).}
\label{fig_10yrs_pair_hist}
\end{figure}

In addition to uniformity of the coverage, the LSST mission calls for pairs of visits within a valid time window at the same night. The main reason is to detect the transient objects such as asteroids. Since, the moving objects usually belong to the solar system, the pair constraint was initially imposed only on the WFD and NES regions. However, there are interesting solar system objects such as interstellar asteroids that can be observed in any direction of the sky, besides identification of the other varying objects, such as super novae, can benefit from a follow up visit, especially if the second visit is with a different filter. Thus in the Modified Feature-Based scheduler we made the pair constraint a universal constraint for all of the regions. The downside of this extension is the fact that it constrains the scheduler even more and the performance can be potentially less than it could be. Note that the structure of the Feature-Based scheduler, allows for extension or restriction of the constraints down to the individual field's level, with neither contradicting any of the Markovian framework assumptions, nor breaking the structure of the implementation. Figure~\ref{fig_10yrs_pair} demonstrates the distribution of the observations in pairs (in the $g$, $r$, and $i$ filters) to the total number of the observations. For the regions that the pair constraint is applied, this ratio can be interpreted as the success rate of the scheduler satisfying the pair constraint. Figure~\ref{fig_10yrs_pair_hist}, compares the distribution of the pairs ratio of the modified Feature-Based scheduler and opsim V4. Note that the peak of the density for Figure~\ref{fig_10yrs_pair_hist} is closer to $1$, which means a larger area of the sky is covered by successful pairs, however, the Modified Feature-Based scheduler offers a sharper concentration of the values that can be interpreted as a more homogenous pairs ratio. In other words, the Modified Feature-Based scheduler, sacrifices perfect pairs observation for a limited area of the sky to maintain a uniform ratio of pairs for a larger area of the sky.

\subsection{AltAz and airmass distributions}
\begin{figure}[h!]
\begin{center}$
\begin{array}{rl cc rl}
\includegraphics[width=.25\linewidth]{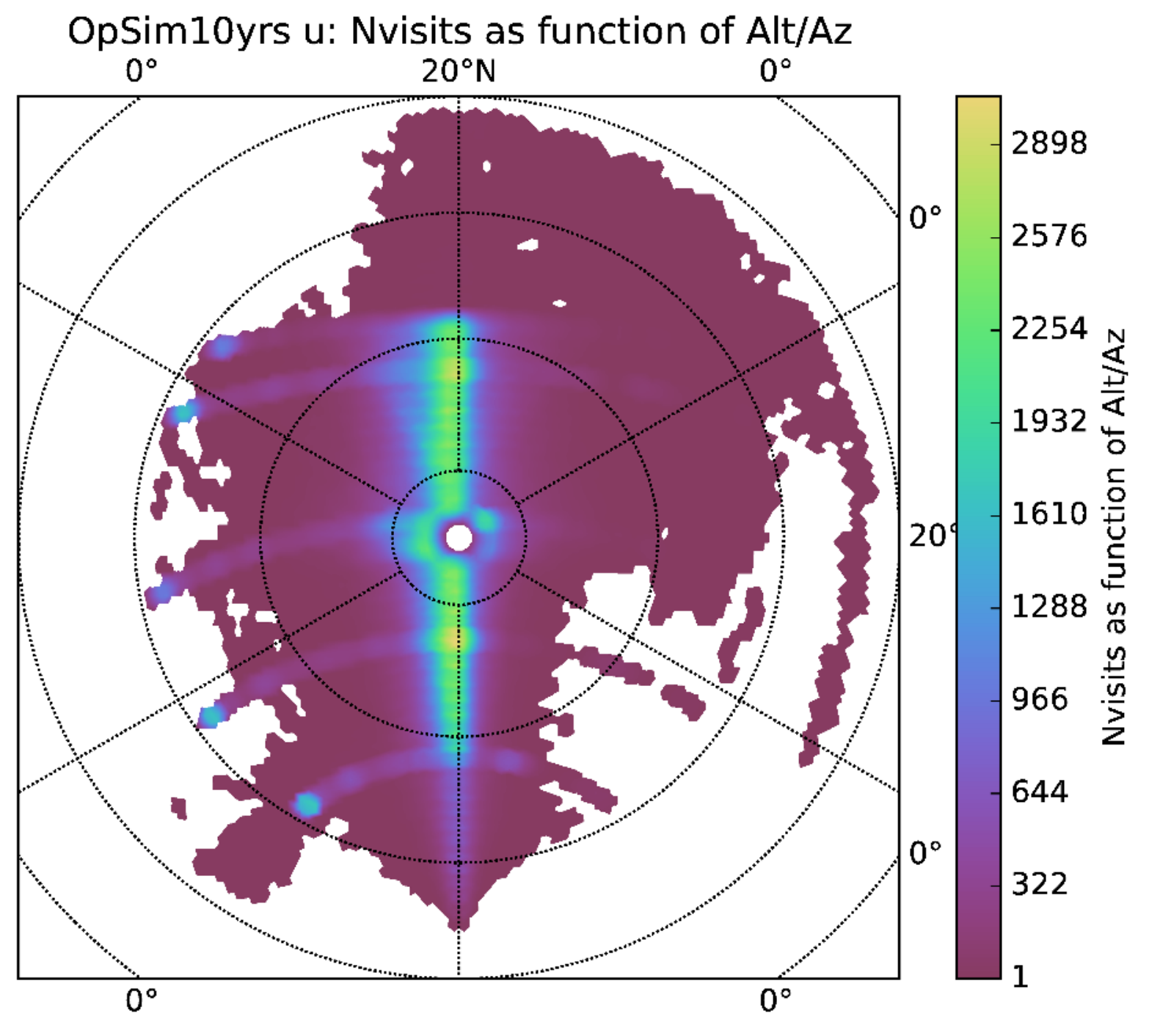}&
\includegraphics[width=.25\linewidth]{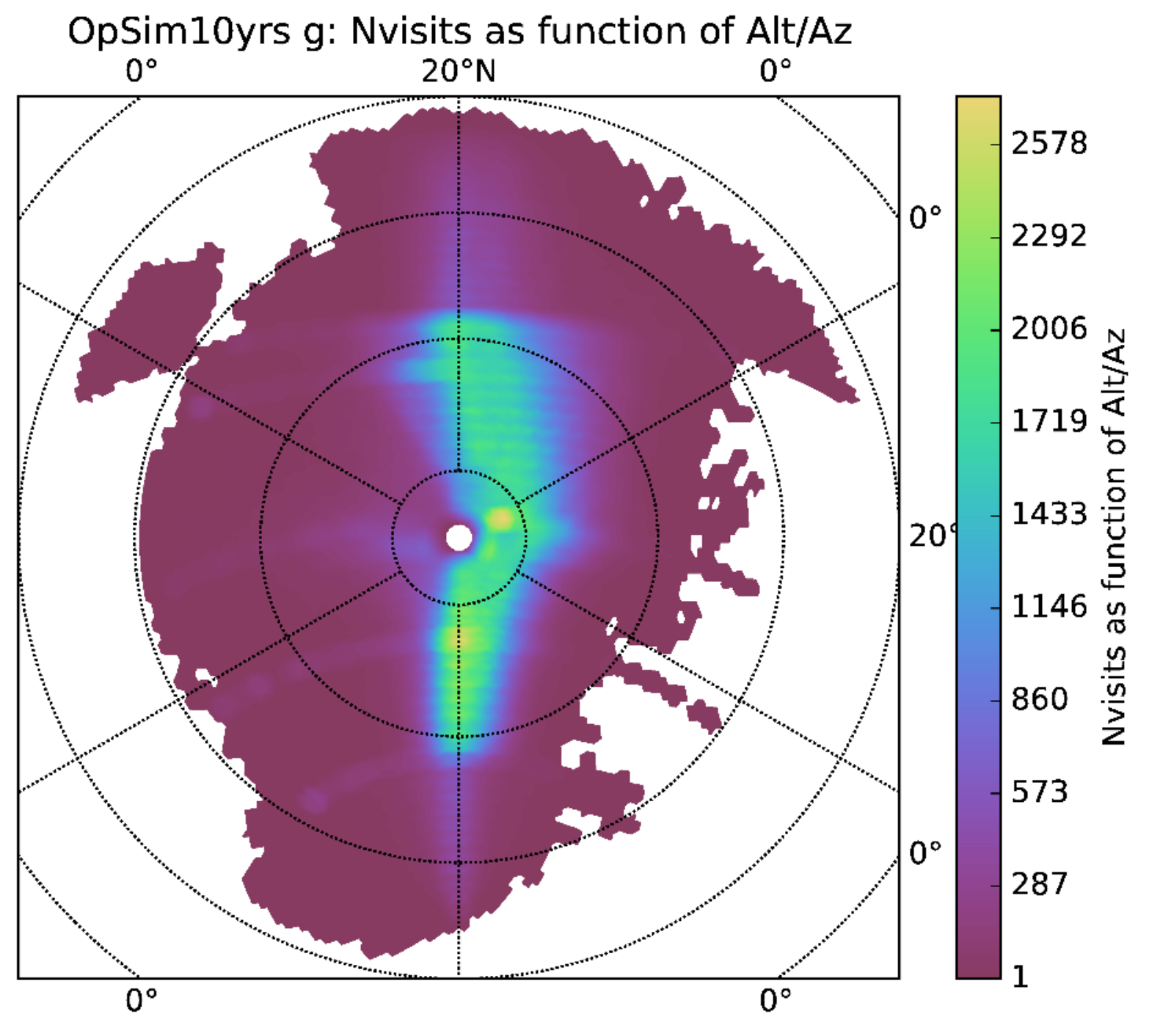}&    &  &
\includegraphics[width=.25\linewidth]{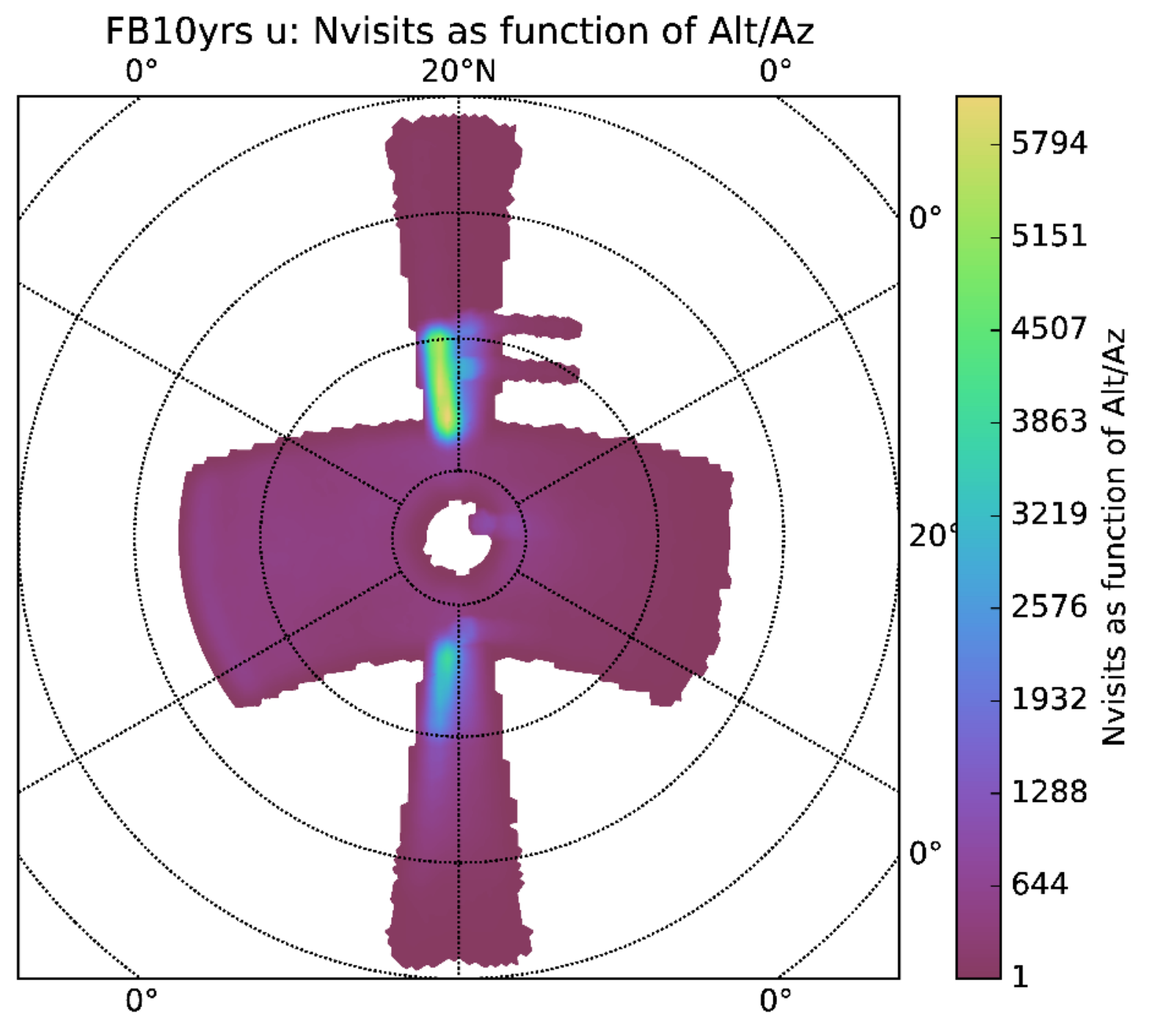}&
\includegraphics[width=.25\linewidth]{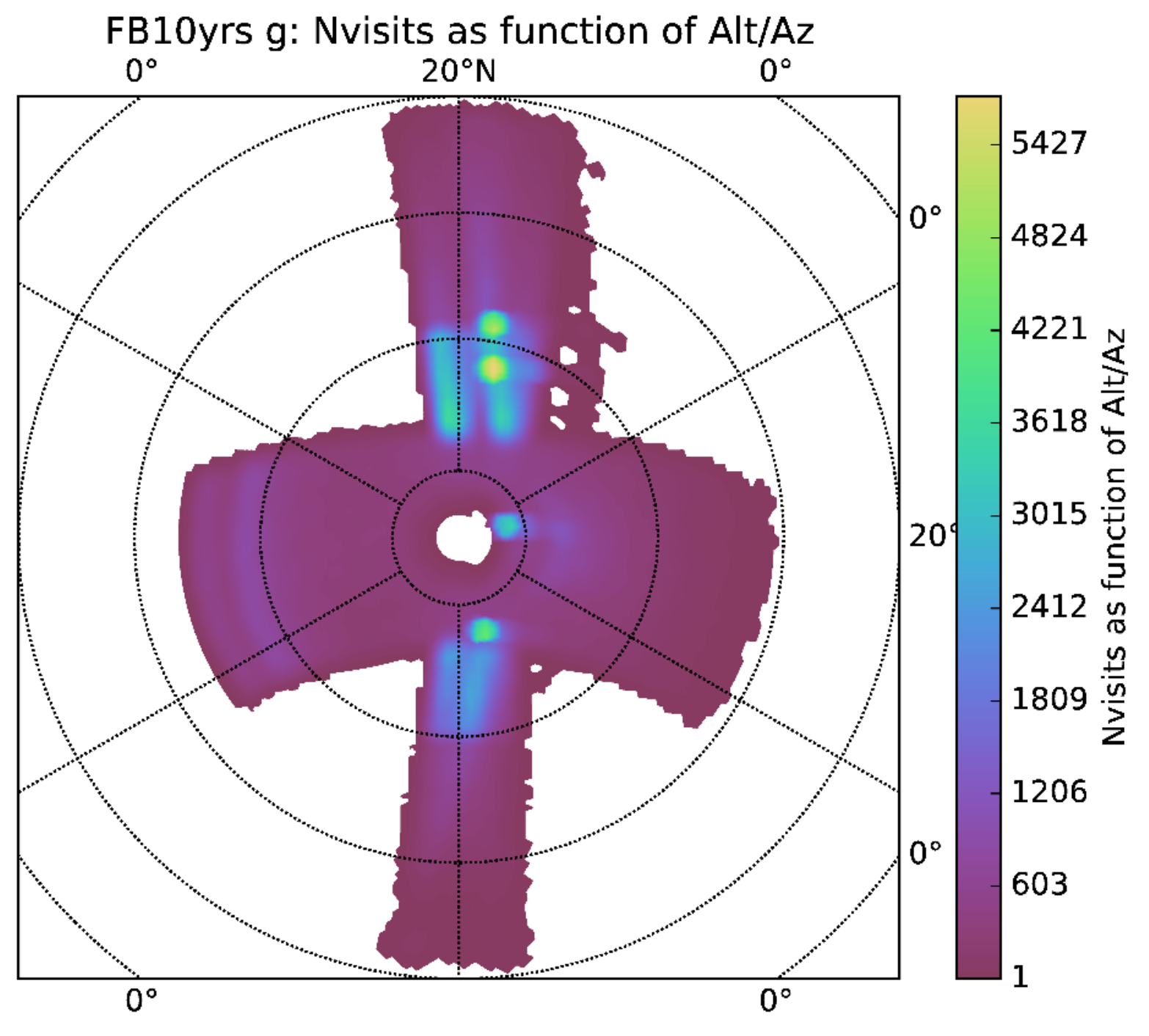}\\
\includegraphics[width=.25\linewidth]{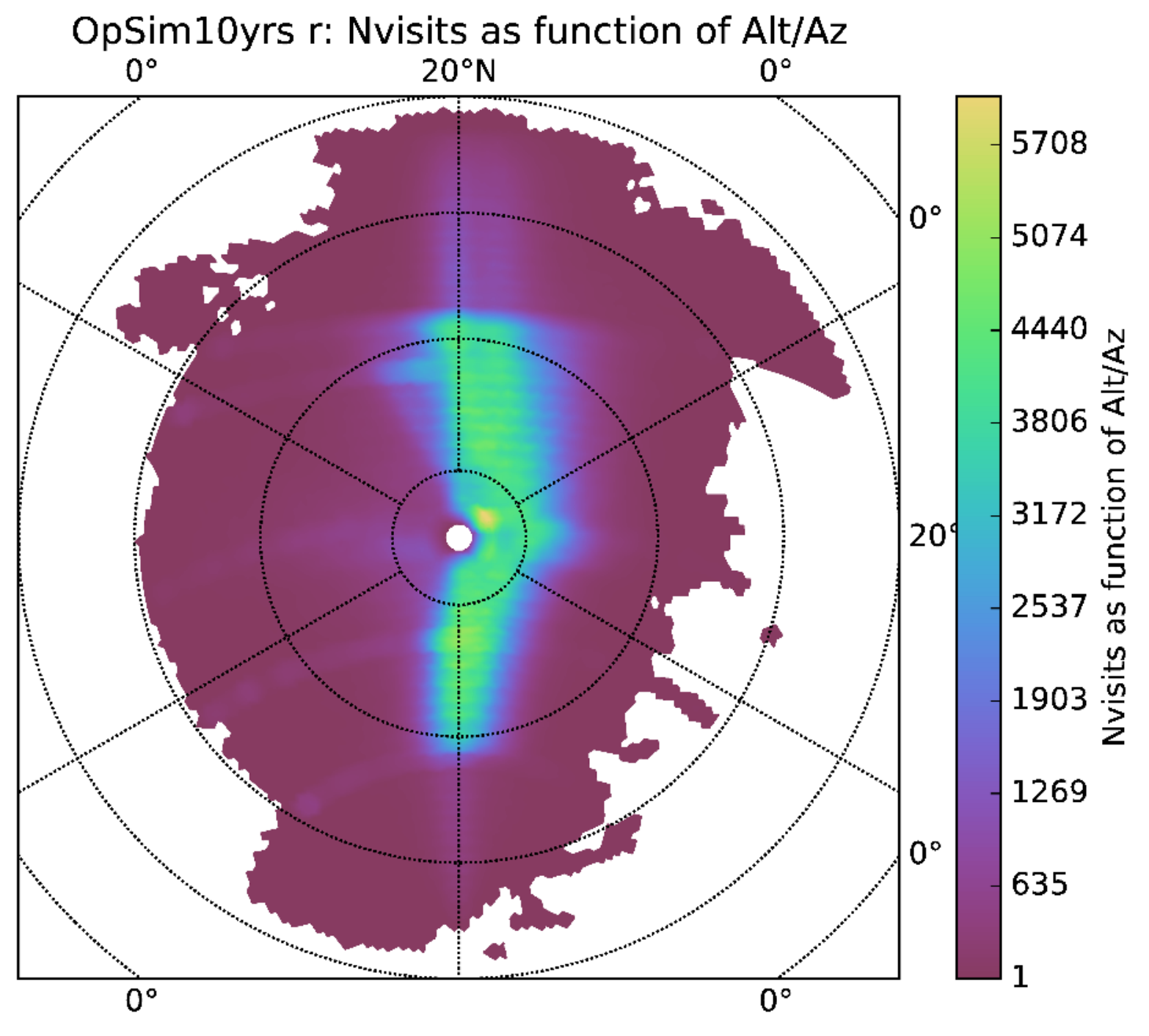}&
\includegraphics[width=.25\linewidth]{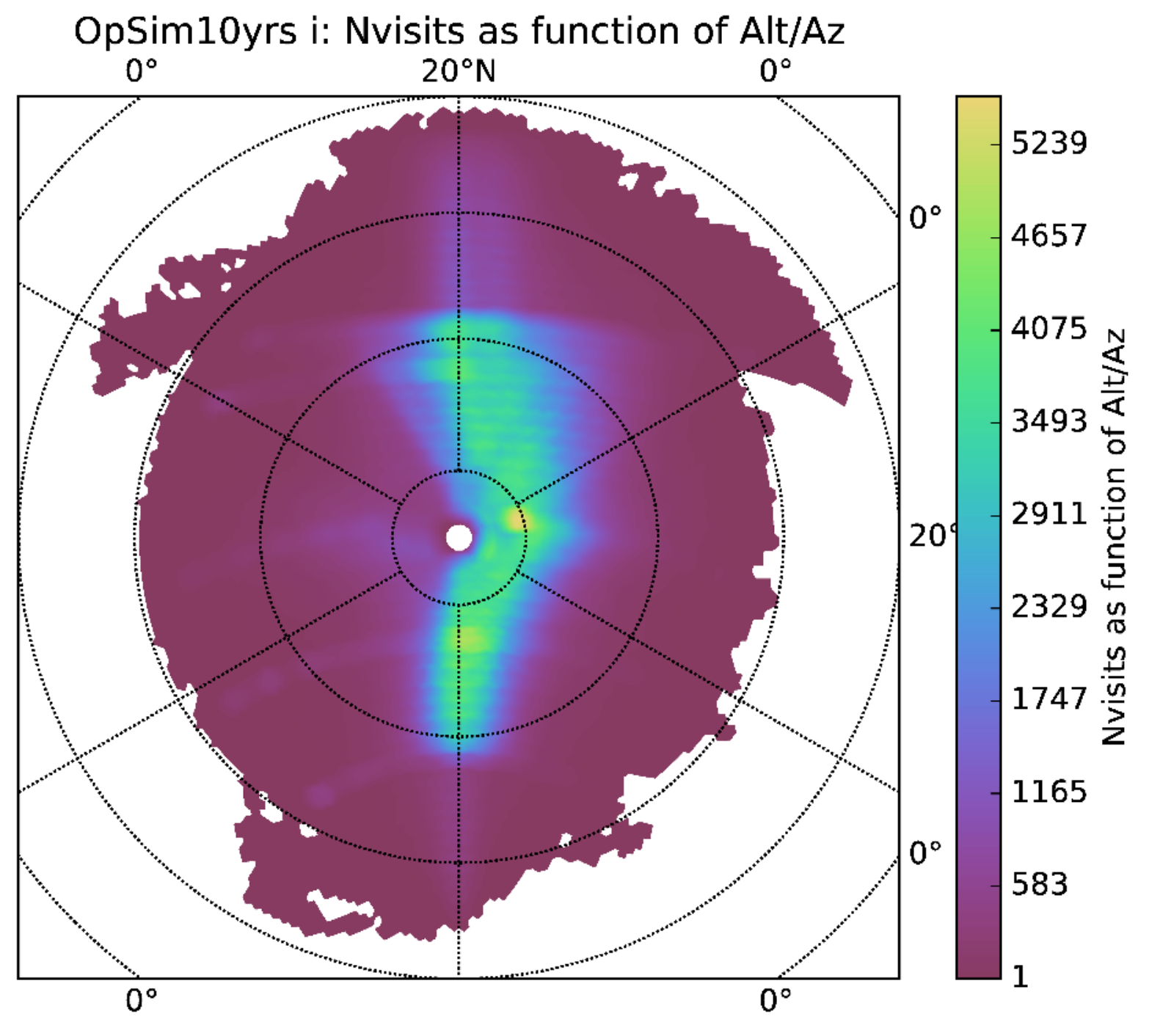}&    &  &
\includegraphics[width=.25\linewidth]{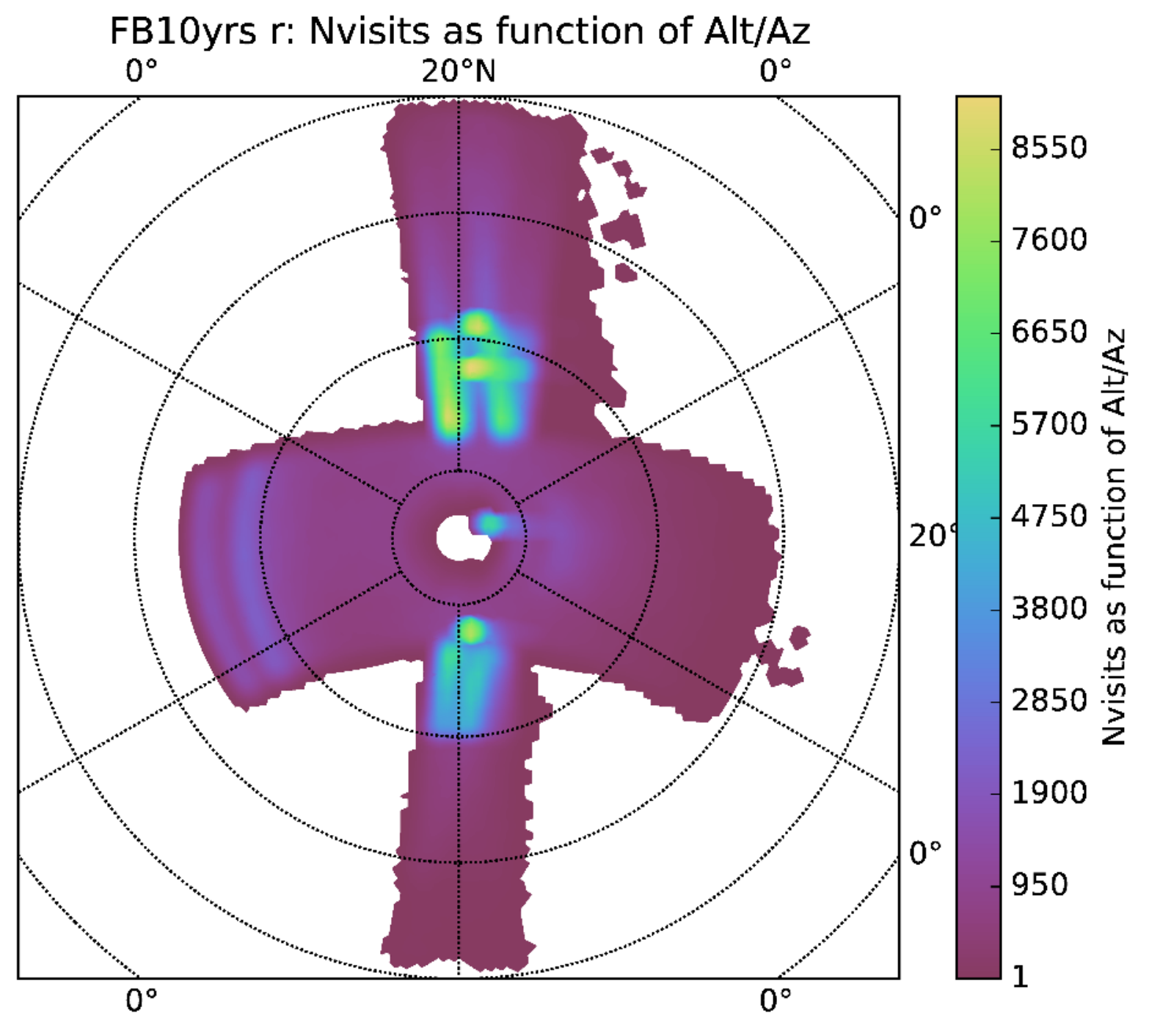}&
\includegraphics[width=.25\linewidth]{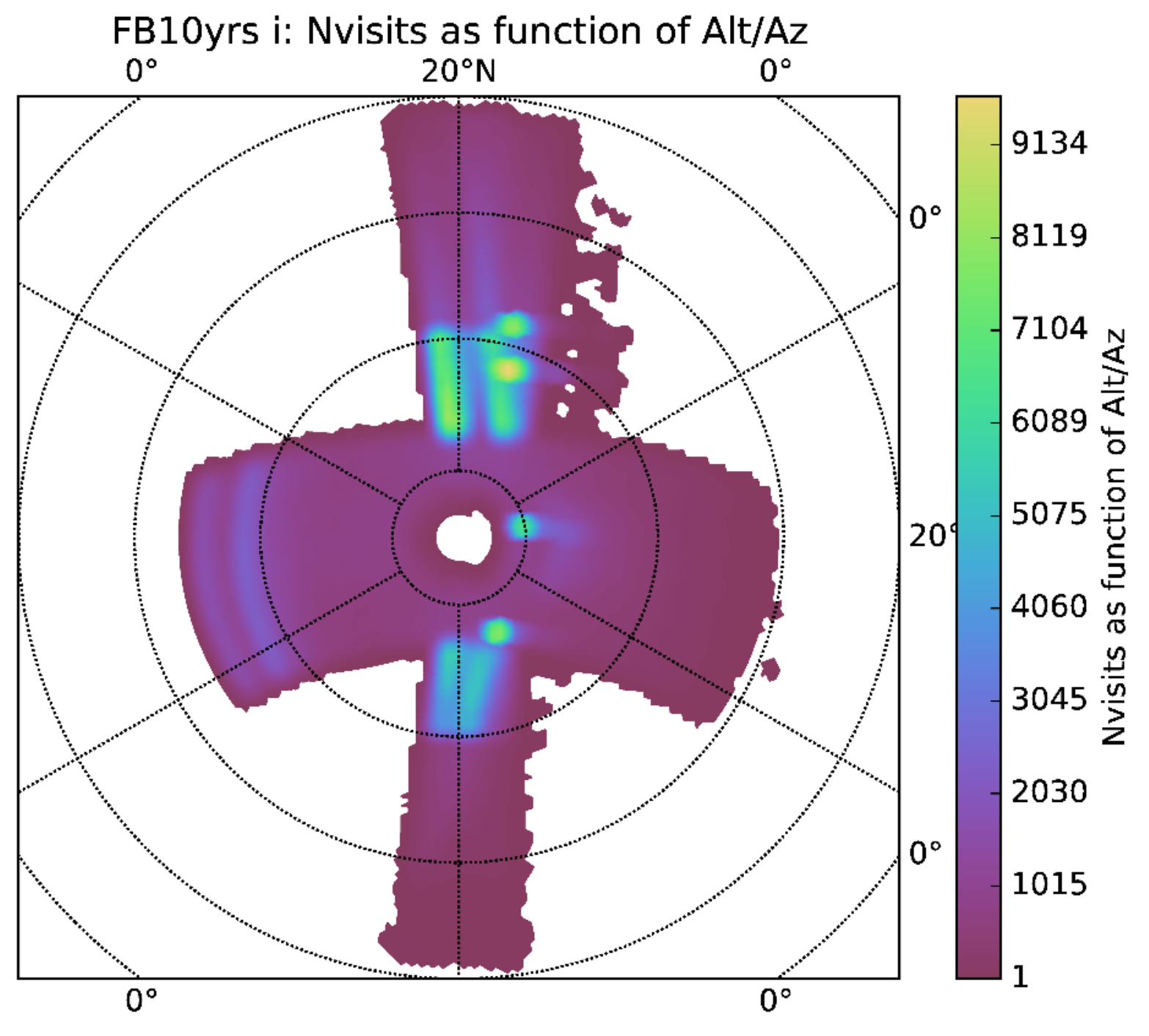}\\
\includegraphics[width=.25\linewidth]{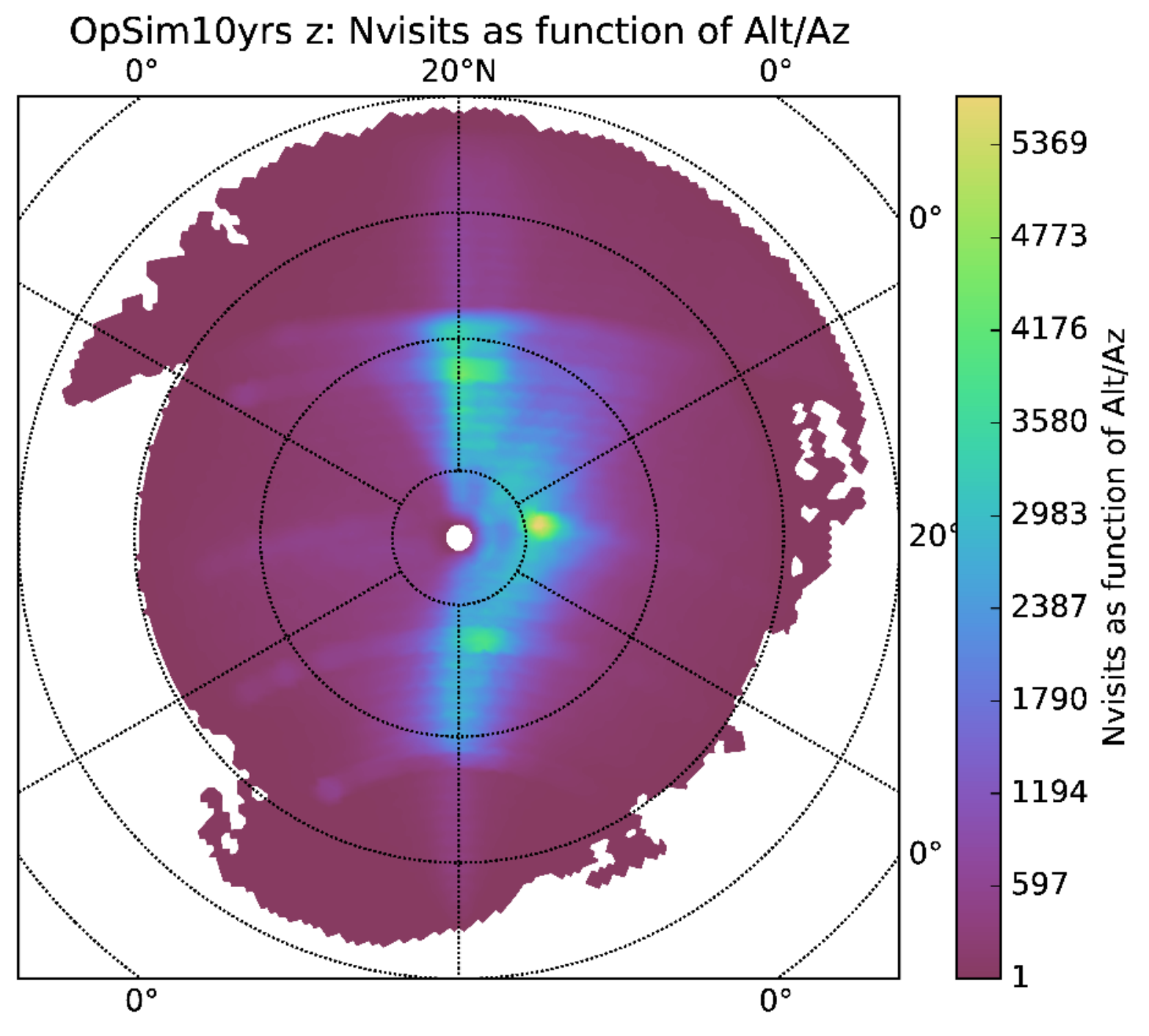}&
\includegraphics[width=.25\linewidth]{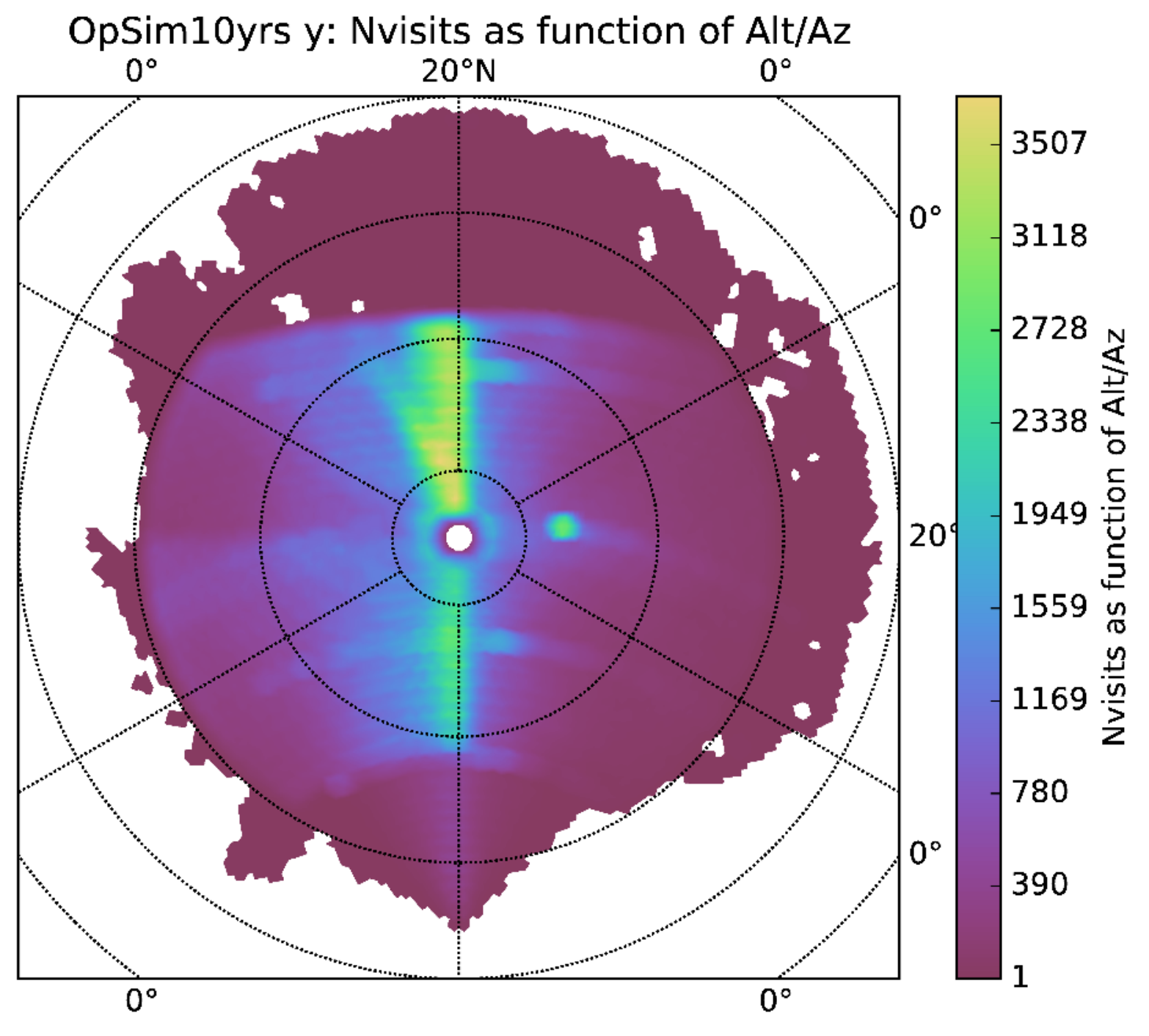}&    &  &
\includegraphics[width=.25\linewidth]{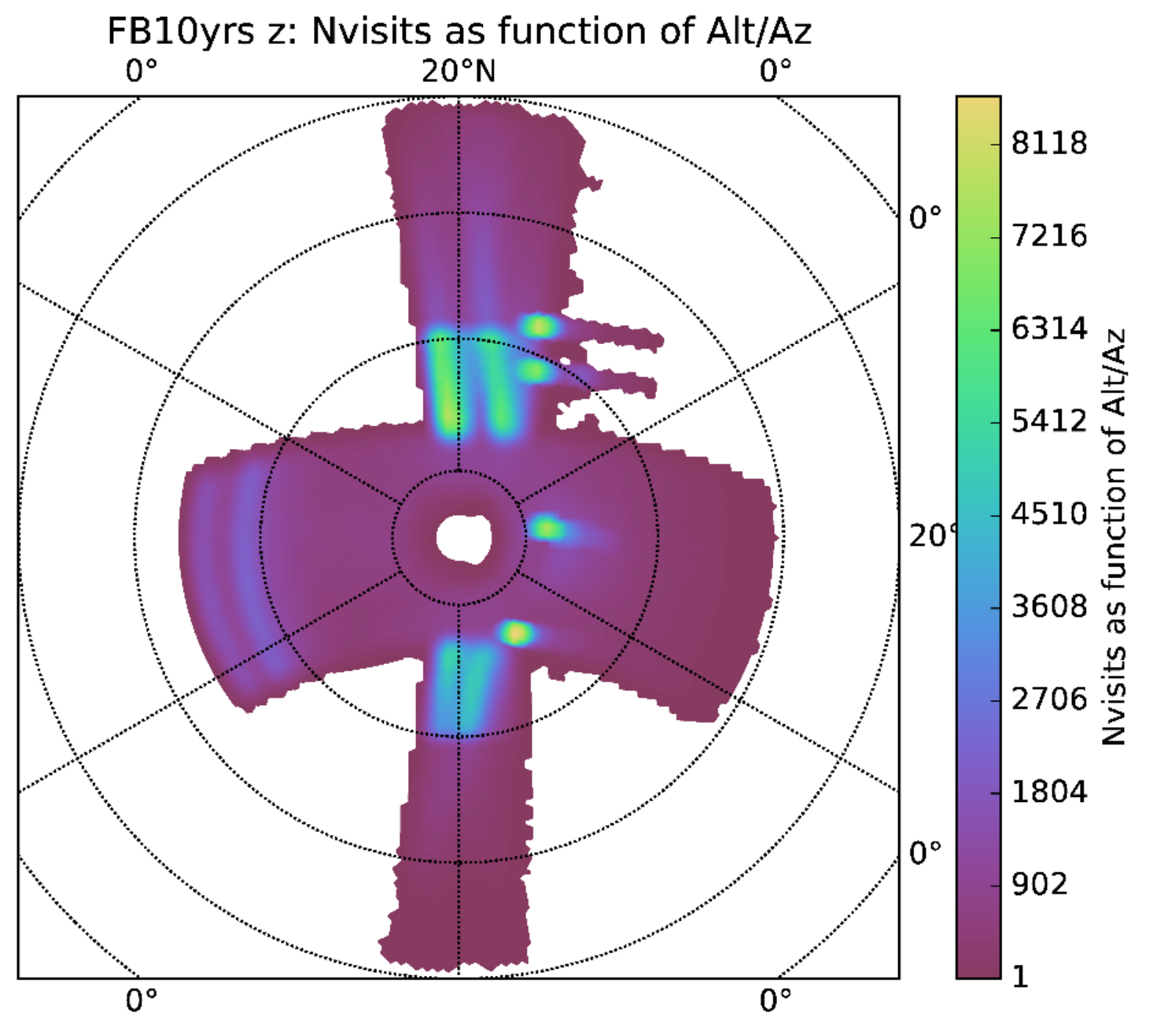}&
\includegraphics[width=.25\linewidth]{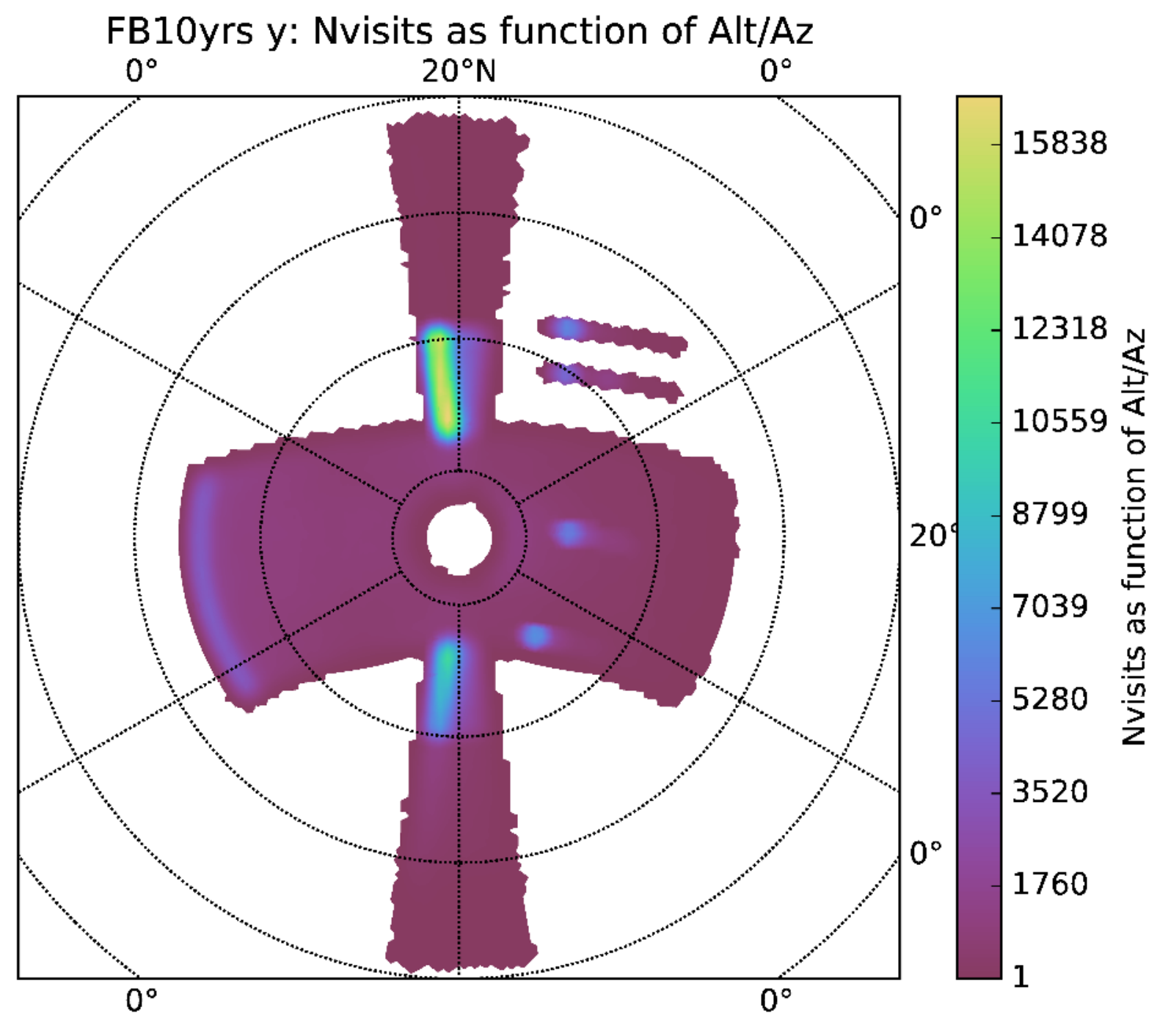}\\
\end{array}$
\end{center} 
\caption{Each plot is the distribution of the visits on an altitude-azimuth sky map in one of the six filters. The two left columns belong to opsim V4, and the two right columns belong to a simulation of the Modified Feature-Based scheduler. The higher concentration on the meridian (vertical axis) for the Modified feature-Based scheduler shows a more desirable behavior. Moreover, consistent concentration of the visits on the east wing can potentially provide a better pairs observation.}
\label{fig_10yrs_AltAz}
\end{figure}

\begin{figure}
\epsscale{0.35}
\plotone{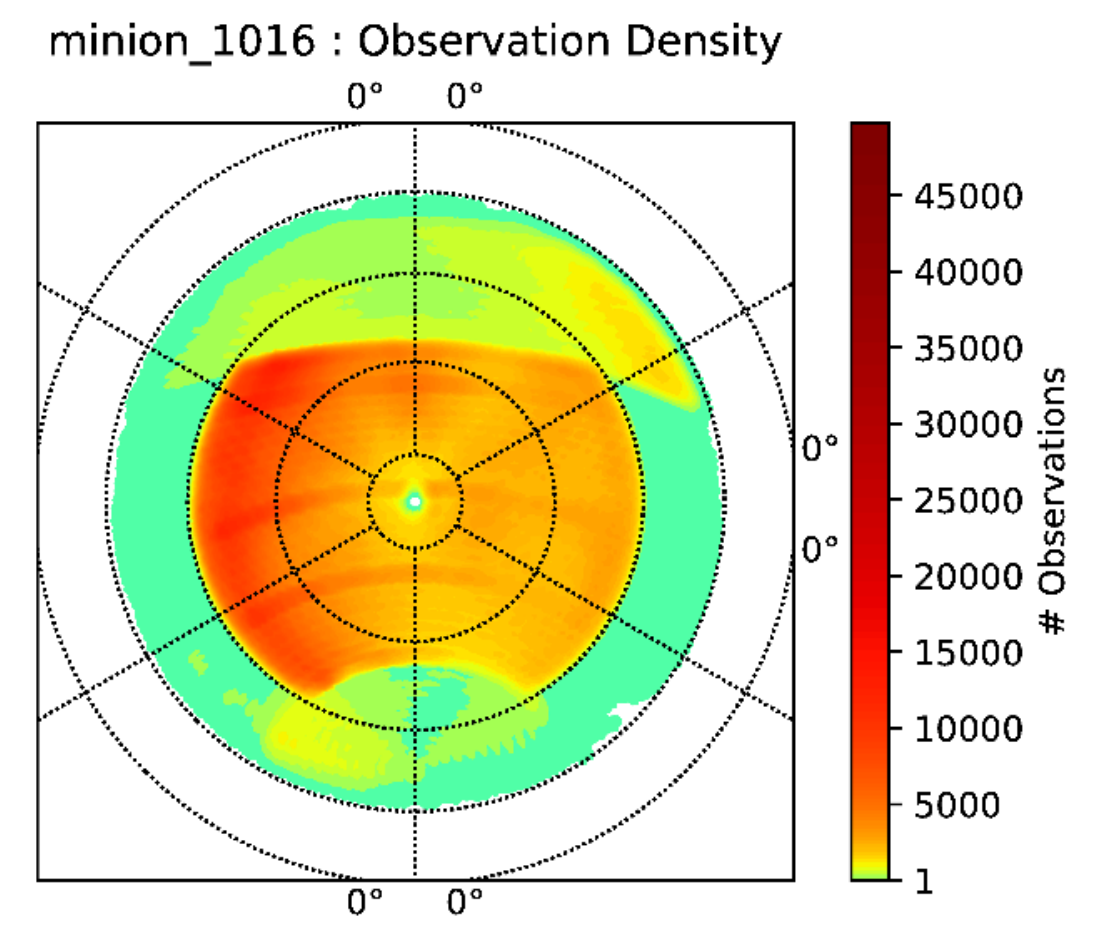}
\plotone{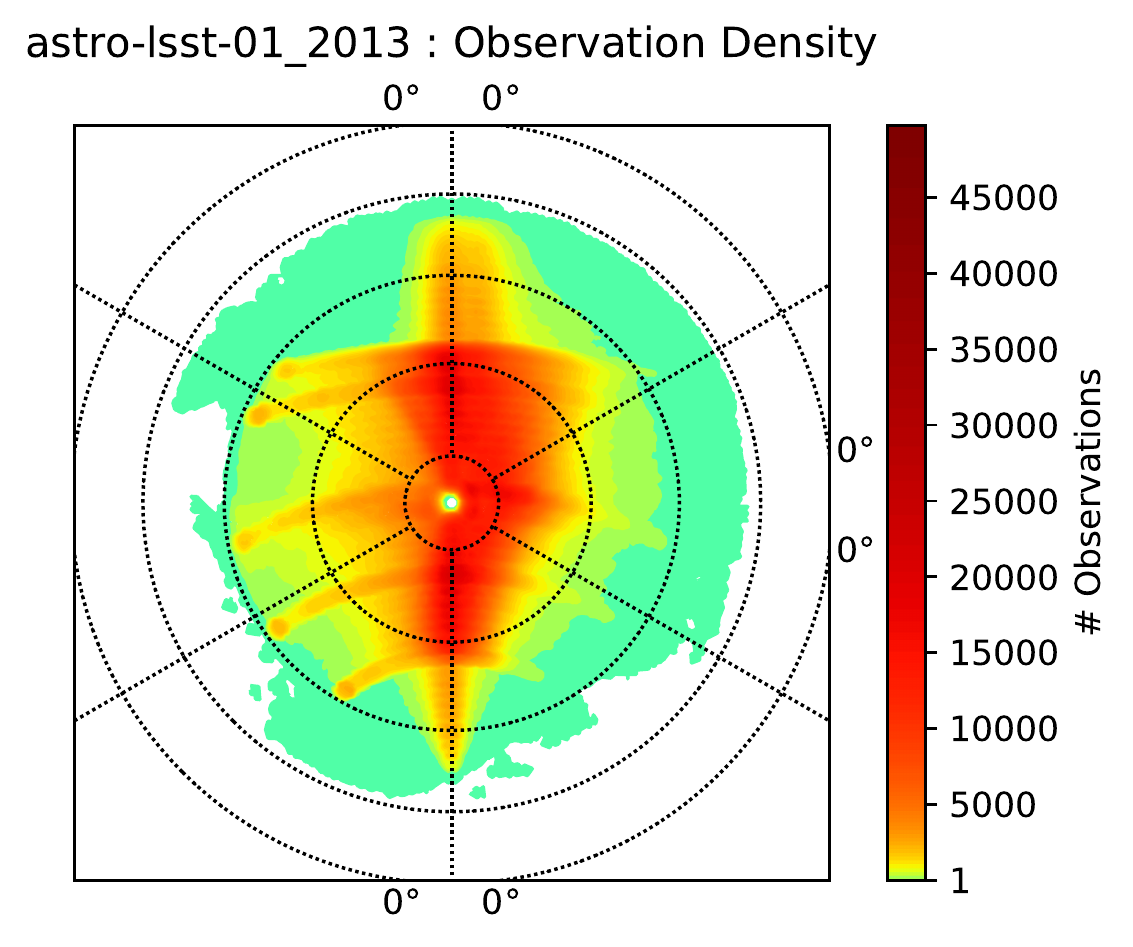}
\plotone{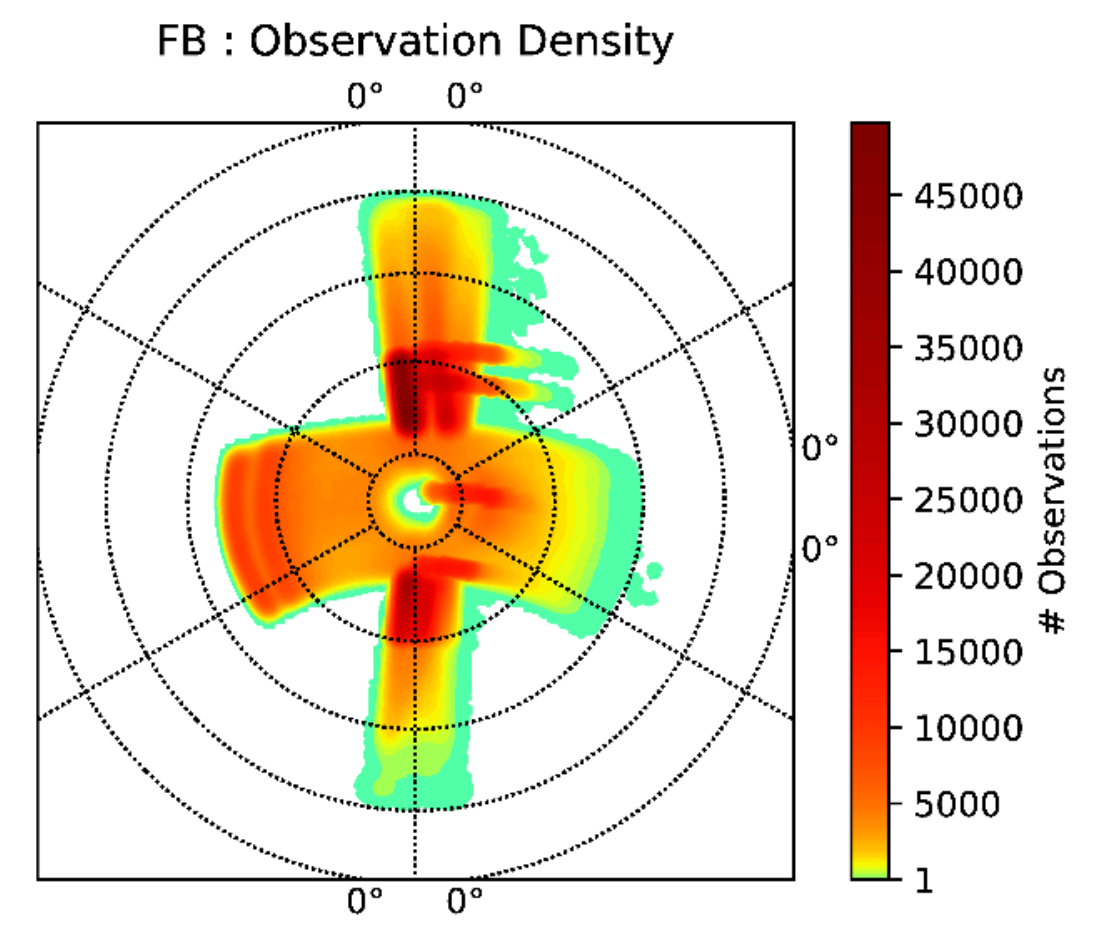}
\epsscale{1}
\caption{The distribution of the visits on an altitude-azimuth sky map for three schedulers. On the left, opsim V3 schedules most observations at high airmass and very few on the meridian. In the middle, opsim V4 schedules many observations on the meridian, but still executes deep drilling fields and a smattering of other observations at high airmass. On the right, the Modified Feature-Based scheduler concentrates observations around the meridian, including deep drilling observation. }\label{fig_all_alt_az}
\end{figure}

For a ground-based instrument, airmass is one of the major obstacles for high-quality observation. Although zenith observations have the minimum airmass, off-zenith observations cannot be avoided, in which case, observations around the meridian provide high quality data and consequently result in more efficient operation of the telescope. Figure~\ref{fig_10yrs_AltAz} compares the number density of the visits on an altitude-azimuth sky map in each of the six filters $[u,g,r,i,z,y]$. Clearly in all of the filters, the modified Feature-Based scheduler schedules more visits around the desirable meridian zone. In addition, it offers a consistent concentration peak on the east wings, which is essential for a higher success rate for the pairs constraint. Because, if the first visit of the night occurs when the field is on the east side of the sky, it provides a longer opportunity for the second visit of the same night. Figure \ref{fig_all_alt_az}, demonstrates the density of visits collectively in all filters for opsim V3 and opsim V4, and the Modified feature-Based scheduler. Note that, adjustability of the Feature-Based scheduler allows for a significant change in the behavior, in this case by defining a new basis function, the Modified Feature-Based scheduler prefers to observe a contiguous set of fields that is then re-observed later in the same order.

\subsection{Signal-to-Noise ratio}
For a multi-objective survey telescope, such as LSST, comparing the overall performance of the different schedules is a difficult task. Particularly because of the large number of the competing factors that are involved in the performance evaluation. In some cases involving the importance of each area of astronomy, the criteria are not even objective. Nevertheless we conclude this section with a general comparison by the median throughput (signal-to-noise ratio) as a general measure for the quality of a schedule. Table~\ref{OSF_table} reflects the value of median throughput for three different schedules in \textit{r} and \textit{g} bands. Modified Feature-Based scheduler significantly outperforms both of the baseline schedules. In addition note that the throughput is mainly determined by the combination of the surveys' open shutter fraction (OSF), and the airmass. The open shutter fraction is the total time that the telescope camera shutter was open divided by the total time it could have been open. This reveals how time-efficiently the observations have been scheduled. The median airmass reflects the overall quality of the collected data. As mentioned before, observations in lower airmass allows for a higher data quality. Comparing the values of OSF and the airmass for both of the baseline schedules, opsim V3 and opsim V4 shows that there is a trade-off between the two values. While opsim V4 offers better median airmass, its OSF is decreased. However, its median throughput is very close to that of opsim V3. This comparison reveals that the change of meta-parameters and objectives in the scheduler of opsim V3 and opsim V4, changes the balance of trade-off between OSF and airmass, but not the actual performance of the scheduling, measured by the median thtoughput.

\begin{table}
\caption{Comparison of three different survey algorithms in a section of the LSST Wide-Fast-Deep survey area.}\label{OSF_table}
\begin{center}
\begin{tabular}{l|ccccc} \hline
Survey &\multicolumn{2}{c}{median throughput} & OSF & median Airmass & dithered\\
&$r$ (\%) &  $g$ (\%) &&&\\ \hline
Modified F-B & 63.7 & 47.0& 0.705 & 1.1 & yes  \\
opsim V3      & 55.3 & 40.0 & 0.736 & 1.2 & no  \\
opsim V4      & 54.4 & 40.8 & 0.715 & 1.1 & no \\ \hline
\end{tabular}
\end{center}
\end{table}

\section{Concluding remarks}\label{sec_conclusion}

This study demonstrated that a Markovian scheduler based on expert-designed features, and a parametrized linear decision making policy can be successfully applied to multi-mission, ground-based telescopes such as LSST. Unlike the mainstream telescope schedulers, Feature-Based scheduler does not rely on hand-crafted observation proposals. Instead, by bringing the decision making process to the individual observation level, improves the efficiency of the telescope's operation. In particular, because with this approach the telescope's schedule can be designed for optimality in addition to feasibility, and in general, because schedulers which rely on human interaction are fundamentally prone to potential suboptimality. This is mainly due to the manual tailoring which is performed based on the inspections of the instances. Moreover, adjusting the behavior of human dependent schedulers are inconvenient and time consuming in practice. Furthermore, being modeled as a Markovian Decision Process, the Feature-Based scheduler offers a systematic approach to the optimization of the scheduler's behavior under uncertainties and interruptions.

On the other hand, the decision elements in the Feature-Based scheduler are designed, and separated in an intuitive way for the astronomy community. This property allows for expert intervention if needed, however in a regulated way, and only on the parameters of the policy. Manual adjustments of the parameters of the policy does not break measurability, linearity, and memorylessness of the process. Thus all of the simplicity, optimality conditions and modularity of the design remain valid. In addition, due to the coherent structure, from training to online decision-making, the Feature-Based scheduler is easy to understand, implement, and troubleshoot. Simplicity of the design and implementation, also provides a desirable environment for a wide-range of programming expertise in astronomy community to install the python packages on a local computer, define a custom mission objective, train a scheduler and examine the behavior of the scheduler with various mission objectives. Similarly, in a particular project, when a change in the mission's objective is necessary, deriving a new scheduler that optimizes the new objective is principally automated. Furthermore, for the mission planning stage of a future instrument, a scheduler with adjustable objective can be extremely helpful, because it can answer the high-level trade-off questions, such as the amount of the time efficiency loss in the different strategies of capturing transient objects.

Computationally, the required resources for the training/optimizing of the Feature-Based scheduler is versatile, depending on the purpose. If many different objective functions are being tested for planning a mission, then a quick $e$DE optimization for a short scheduling episode can find a sufficiently good scheduler for each mission. Even a quick manual hand tuning that reflects the intuitive importance of each basis function is possible, because they carry an astronomical observation meaning. On the other hand if the objective is known and fixed, and the scheduler is being trained for real-time decision making, then one might even categorize the observation nights based on their main differences, such as the moon-phase, seasonal variations, and weather patterns then train a scheduler specifically for each category to further increase the efficiency of the scheduler. 

\appendix

\begin{proof} (of Proposition \ref{prop_main})
Consider a function $C_{\pi}: \pazocal{S} \rightarrow \rm I\!R$, defined as follows,
\begin{equation*}
\begin{aligned}
C_{\pi}(x_{i}) &= -E_{\pi}[ \sum_{i\leq j}^N \gamma^{i - j} R_{\pi(X_j)}(X_{j}, X_{{j+1}}) |x_{i}]\\
& = - E_{\pi}[R_{\pi(x_i)}(x_{i}, X_{{i+1}}) | x_{i}] - \gamma E_{\pi}[ \sum_{ i+1 \leq j}^N \gamma^{j-(i+1)} R_{\pi(X_j)}(X_{j}, X_{j+1}) | x_{i}].\\
\end{aligned}
\end{equation*}
Then by applying the the law of total expectation on the second term,
\begin{equation*}
\begin{aligned}
C_{\pi}(x_{{i}}) &= - E_{\pi}[R_{\pi(x_i)}(x_{i}, X_{{i+1}}) | x_{i}] - \gamma E_{\pi}[ E_{\pi}[ \sum_{ i+1 \leq j}^N R_{\pi(X_j)}(X_{j}, X_{{j+1}}) |X_{{i+1}}]| x_{i}]\\
&=  - E_{\pi}[R_{\pi(x_i)}(x_{i}, X_{{i+1}}) | x_{i}] - \gamma E_{\pi}[ C_{\pi}({X_{{i+1}}})| x_{i}],
\end{aligned}
\end{equation*}
and by assuming a finite state space,
\begin{equation*}
C_{\pi}(x_{{i}}) =  - E_{\pi}[R_{\pi(x_i)}(x_{i}, X_{{i+1}}) | x_{i}]  - \gamma \sum_{x_{i+1} \in \pazocal{S}} \mathbb P(x_{i+1}|\pi(x_i), x_{i}, x_{i-1},\dots, x_{0}) C_{\pi}({x_{{i+1}}}).\\
\end{equation*}
Then by the Markov property,
\begin{equation*}
\begin{aligned}
C_{\pi}(x_{i}) &=  - E_{\pi}[R_{\pi(x_i)}(x_{i}, X_{{i+1}}) | x_{i}] - \gamma \sum_{x_{i+1} \in \pazocal{S}} P_{\pi(x_{i})}(x_{i},x_{i+1}) C_{\pi}({x_{{i+1}}}),\\
\end{aligned}
\end{equation*}
where, $P_{\pi(x_{i})}(x_{i},x_{i+1}) $ is the transition probability from $x_i$ to $x_{i+1}$ under the action of $\pi(x_i)$. 
Now, let $C^*(x_{i}) = \min_{\pi} C_{\pi}(x_{{i}})$ then,
\begin{equation}\label{equ_proof1}
\begin{aligned}
C^{*}(x_{{i}}) &=  ~~~~~~~ \min_{\pi} ( - E_{\pi}[R_{\pi(x_i)}(x_{i}, X_{{i+1}}) | x_{i}] - \gamma \sum_{x_{i+1} \in \pazocal{S}} P_{\pi(x_{i})}(x_{i},x_{i+1}) C_{\pi}({x_{{i+1}}}) ) \\
& =  \min_{\{\pi(x_{i}), \pi(x_{{i+1}}),\dots \}}(- E_{\pi}[R_{\pi(x_i)}(x_{i}, X_{{i+1}}) | x_{i}] - \gamma \sum_{x_{i+1} \in \pazocal{S}} P_{\pi(x_{i})}(x_{i},x_{i+1}) C_{\pi}({x_{{i+1}}}))\\
& = ~~~~~~~ \min_{\pi(x_{i})} (- E_{\pi}[R_{\pi(x_i)}(x_{i}, X_{{i+1}}) | x_{i}] - \min_{\{\pi(x_{i+1}), \pi(x_{{i+2}}),\dots \}} \gamma \sum_{x_{i+1} \in \pazocal{S}} P_{\pi(x_{i})}(x_{i+1},x_{i}) C_{\pi}({x_{{i+1}}})).\\
\end{aligned}
\end{equation}
Note that both of the one step reward $R_{\pi(x_i)}(x_{i}, X_{{i+1}})$ and transition probability $P_{\pi(x_{i})}(x_{i}, X_{{i+1}})$ depend only on $\pi(x_{i})$ which is the action taken at $t_i$. 
For the next time step, one can construct a $\hat {C}$ function such that, $\hat C(x_{{i+1}}) = \min_{\pi(x_{{i+1}})} \min_{\pi} C_{\pi} ({x_{i+1}})$, then, 

\begin{equation*}
\begin{aligned}
\sum_{x_{i+1} \in \pazocal{S}} P_{\pi(x_{i})}(x_{i},x_{i+1}) \min_{\{\pi(x_{i+1}), \pi(x_{{i+2}}),\dots \}} C_{\pi}({x_{{i+1}}})&= \sum_{x_{i+1} \in \pazocal{S}} P_{\pi(x_{i})}(x_{i},x_{i+1})  \hat C ({x_{{i+1}}} )\\
& \geq \min_{\pi} \sum_{x_{i+1} \in \pazocal{S}} P_{\pi(x_{i})}(x_{i+1},x_{i}) C_{\pi}({x_{{i+1}}}).
\end{aligned}
\end{equation*}
On the other hand, 
\begin{equation*}
\sum_{x_{i+1} \in \pazocal{S}} P_{\pi(x_{i})}(x_{i},x_{i+1}) \min_{\{\pi(x_{i+1}), \pi(x_{{i+2}}),\dots \}} C_{\pi}({x_{{i+1}}}) \leq \min_{\pi} \sum_{x_{i+1} \in \pazocal{S}} P_{\pi(x_{i})}(x_{i},x_{i+1}) C_{\pi}({x_{{i+1}}}).
\end{equation*}
Therefore,
\begin{equation*}
 \min_{\pi} \sum_{x_{i+1} \in \pazocal{S}} P_{\pi(x_{i})}(x_{i},x_{i+1}) C_{\pi}({x_{{i+1}}})=\sum_{x_{i+1} \in \pazocal{S}} P_{\pi(x_{i})}(x_{i},x_{i+1}) \min_{\{\pi(x_{i+1}), \pi(x_{{i+2}}),\dots \}} C_{\pi}({x_{{i+1}}}).
\end{equation*}
By substituting the second term of the right hand side of Equation (\ref{equ_proof1}) with the right hand side of the above equation,
\begin{equation*}
\begin{aligned}
C^{*}(x_{{i}}) &= \min_{\pi(x_{i})} (- E_{\pi}[R_{\pi(x_i)}(x_{i}, X_{{i+1}}) | x_{i}] -\gamma \sum_{x_{i+1} \in \pazocal{S}} P_{\pi(x_i)}(x_{i},x_{i+1}) \min_{\pi} C_{\pi}({x_{{i+1}}}))\\
&= \min_{\pi(x_{i})} (- E_{\pi}[R_{\pi(x_i)}(x_{i}, X_{{i+1}}) | x_{i}] -\gamma \sum_{x_{i+1} \in \pazocal{S}} P_{\pi(x_i)}(x_{i},x_{i+1}) C^*({x_{{i+1}}})).\\
\end{aligned}
\end{equation*}
The last equality follows from the definition of $C^*$, and is in the form of Optimal Bellman Equation, for which a solution exists [Bellman 1957], and $C^*(x_0) = \min{C^\pi(x_0)}$ attains the optimal value, by the construction of $C^\pi(x_0)$ which is equal to $- E[\sum _{\pi(x_i)}(x_{{i}}, X_{i+1}) | x_0]$.

Given a $C^*$, an optimal policy can be simply evaluated,
\begin{equation}\label{equ_proof2}
\begin{aligned}
\pi^*(x_{i}) & = \argmin_{a_{i} \in \pazocal{A}_i} (- E_{\pi}[R_{a_i}(x_{{i}}, X_{i+1}) | x_{i}] -\gamma \sum_{x_{i+1} \in \pazocal{S}} P_{a_i}(x_{i},x_{i+1})C^*({x_{{i+1}}}))\\
& = \argmin_{a_{i} \in \pazocal{A}_i} (- E_{\pi}[R_{a_i}(x_{{i}}, X_{i+1}) - \gamma C^*({X_{{i+1}}})| x_{i}]).\\
\end{aligned}
\end{equation}
Finally, define,
\begin{equation}\label{equ_phi}
\Phi(X_{{i+1}}) := R_{a_i}(x_{{i}}, X_{i+1}) - \gamma C^*({X_{{i+1}}}),
\end{equation}
 and substitute in Equation (\ref{equ_proof2}) to complete the proof.
\end{proof}

\acknowledgments
\textbf{Acknowledgments.} We would like to thank the LSST team, and the DIRAC Institute's faculty and researchers for providing expertise, without which this research would not have been possible. In particular, Professor \v{Z}eljko Ivezi\'{c} whose consistent attention and insightful comments greatly assisted this work. We would also like to show our gratitude to the experts in other institutes, specially Professor Michael Strauss, Professor Christopher Stubbs, Dr. Robert Lupton, and Dr. Michael Reuter for their insightful comments spanning from the idea stage to the end of this project. This work is financially supported by the National Science Foundation under Cooperative Agreement 1258333 managed by the Association of Universities for Research in Astronomy (AURA), and the Department of Energy under Contract No. DE-AC02-76SF00515 with the SLAC National Accelerator Laboratory. Additional LSST funding comes from private donations, grants to universities, and in-kind support from LSSTC Institutional Members. The DIRAC Institute is supported through generous gifts from the Charles and Lisa Simonyi Fund for Arts and Sciences, and the Washington Research Foundation.

\facility{LSST}

\software{MAF \citep{jones2014lsst}, healpy \citep{Healpy05}, matplotlib \citep{matplotlib07}, 
          astropy \citep{astropy18}, numpy/scipy \citep{scipy}}

\bibliography{references}
\end{document}